\documentclass{article}

\usepackage{tikz}
\usetikzlibrary{arrows,backgrounds}
\usepackage[latin1]{inputenc}
\usepackage{hyperref}
\usepackage{amsmath}
\usepackage{amssymb,amsfonts,amsxtra,amsthm}
\usepackage{stmaryrd}
\usepackage[mathcal,mathscr]{euscript}
\usepackage[english]{babel}
\usepackage{times}

\usepackage[a4paper]{geometry}

\newtheorem{theorem}{Theorem}[section]
\newtheorem{lemma}[theorem]{Lemma}

\newtheorem{corollary}[theorem]{Corollary}
\newtheorem{example}[theorem]{Example} 
\newtheorem{definition}[theorem]{Definition}
\newtheorem{remark}[theorem]{Remark}

\def\ok#1{\mbox{\raisebox{0ex}[1ex][1ex]{$#1$}}}
\def \tuple#1{\langle #1 \rangle}

\newcommand{\cA}{\mathcal{A}}
\newcommand{\cG}{\mathcal{G}}
\newcommand{\cK}{\mathcal{K}}

\newcommand{\ud}{\triangleq}

\newcommand{\Lra}{\Leftrightarrow}
\newcommand{\Ra}{\Rightarrow}
\newcommand{\La}{\Leftarrow}
\newcommand{\comp}{\circ}
\newcommand{\ra}{\rightarrow}
\newcommand{\sra}{\shortrightarrow}

\newcommand{\sq}{\ensuremath{\mathit{sq}}}
\DeclareMathOperator{\uco}{uco}
\DeclareMathOperator{\img}{img}
\DeclareMathOperator{\pre}{pre}
\DeclareMathOperator{\post}{post}
\DeclareMathOperator{\Part}{Part}
\DeclareMathOperator{\Clv}{Cl_\wedge}
\DeclareMathOperator{\Cl}{Cl}
\DeclareMathOperator{\Abs}{Abs}
\DeclareMathOperator{\Sign}{Sign}

\DeclareMathOperator{\States}{States}
\DeclareMathOperator{\paths}{paths}
\DeclareMathOperator{\length}{length}
\DeclareMathOperator{\spu}{sp}

\newcommand{\comment}[1]{}

\newcommand*{\rarr}[1]{\mbox{\raisebox{0ex}[1ex][1ex]{$
  \mathrel{\mathop{
\hspace*{1pt}\longrightarrow\hspace*{1pt}}\limits^{\,_{\mbox{\tiny
\hspace*{-2.2pt}#1}}}}$}}}

\title{Correctness Kernels of Abstract Interpretations}
\author{{\normalsize{\sc Roberto Giacobazzi~~~~~Francesco Ranzato}}\\[10pt]
{\small University of Verona, Italy~~~~~University of Padova, Italy}}

\date{}

\begin{document}

\maketitle

\begin{abstract}
In abstract interpretation-based static analysis, 
approximation is encoded by abstract domains. They
provide systematic 
guidelines for designing abstract semantic functions that approximate some
concrete system behaviors under analysis.  
It may  happen that an abstract domain contains redundant information
for the specific purpose of approximating a given concrete semantic function.
This paper introduces the notion of correctness kernel of abstract interpretations, 
a methodology for simplifying abstract domains, i.e.\
removing abstract values from them, in a maximal way while retaining exactly
the same approximate behavior of the system under analysis. We show that in abstract model checking
correctness kernels provide a 
simplification paradigm of the abstract state space  
that is guided by examples, meaning that this simplification 
preserves spuriousness of examples (i.e., abstract paths). In particular, we show how 
correctness kernels can be integrated with the well-known CEGAR (CounterExample-Guided 
Abstraction Refinement) methodology.
\end{abstract}	

\section{Introduction}\label{intro}

In static analysis and verification, 
model-driven \emph{abstraction refinement} has emerged in the last decade as
a key paradigm for enhancing abstractions towards more precise
yet efficient analyses. 
The underlying basic idea is simple: given an abstraction $A$ modeling some
approximate properties of a system to analyze, 
in order to remove some artificial computations that may
arise in the analysis based on $A$
refine $A$ by considering how the concrete
model actually behaves when false alarms or spurious traces are
encountered. The general idea of using spurious counterexamples for refining an
abstraction stems from the CounterExample-Guided Abstraction Refinement
(CEGAR) paradigm~\cite{cgjlv00,cgjlv03}. The concrete model here drives
the automatic identification of prefixes of the
counterexample abstract path that do not correspond to an actual trace, 
by isolating abstract (failure) states that need to be
refined in order to eliminate that spurious counterexample. Model-driven 
refinement strategies, such as CEGAR, provide algorithmic methods
for achieving abstractions that are complete (i.e., precise \cite{gq01,grs00}) 
with respect to some given property of the concrete
model.

We investigate here the dual problem of 
\emph{abstraction simplification}.  
Instead of refining abstractions in order to eliminate 
spurious traces, our goal is to modify an abstraction $A$
towards a simpler (ideally, the simplest) 
model $A_s$ that gives rise to the same approximate
system behavior as $A$ does. In abstract model checking, this abstraction simplification   
has \emph{to keep the same examples} of the concrete system in the following
sense. Recall that an abstract
path $\pi$ in an abstract transition system $\cA$ is \emph{spurious} 
when no real concrete path is abstracted
to $\pi$. Assume that a 
given abstract state space $A$ of a system $\cA$ 
gets simplified to $A_s$ and thus gives rise to a more abstract 
system $\cA_s$. Then, 
we say that $\cA_s$ keeps the same examples of $\cA$ when
the following condition is satisfied:
if $\pi_{A_s}$ is a spurious path 
in the simplified abstract system $\cA_s$ 
then there exists a spurious path 
$\pi_A$ in the original 
system $\cA$ which is abstracted to $\pi_{A_s}$.   
Obviously, if $\cA_s$ is a generic simplification of $\cA$ then $\cA_s$ does not necessarily 
keep the same examples of $\cA$. In the following, we depict abstract transition
systems by diagrams where integer numbers denote
concrete states, arrows
connect concrete states in a transition relation, and oval shapes indicate blocks 
(denoted by square brackets)
of a state partition. In the example below, 
for the spurious path $\pi_{\cA_s}=\langle[1],[2,3,4],[5]\rangle$ 
in $\cA_s$ there is no corresponding spurious path in $\cA$ which can be abstracted to $\pi_{\cA_s}$
and therefore the simplification $\cA_s$ does not keep the same examples of $\cA$. 

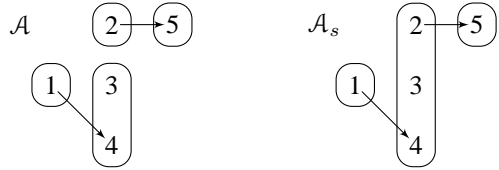
\begin{center}
    \begin{tikzpicture}[scale=0.40]
      \tikzstyle{arrow}=[->,>=latex']
      \path       
      (0,3) node[name=1]{1}
      (2,5) node[name=2]{2}
      (2,3) node[name=3]{3}
      (2,1) node[name=4]{4}
      (4,5) node[name=5]{5};
      \path (-1,5) node[name=a]{$\mathcal{A}$};
      \draw[arrow,shorten >=-5pt, shorten <=-5pt] (1) to (4);
      \draw[arrow,shorten >=-3pt, shorten <=-3pt] (2) to (5);
                        
     \path (1.north west) ++(-0.1,0.1) node[name=a1]{} (1.south east) ++(0.1,-0.1) 
      node[name=a2]{};
     \draw[rounded corners=6pt] (a1) rectangle (a2);

\path (2.north west) ++(-0.1,0.1) node[name=b1]{} (2.south east) ++(0.1,-0.1) 
      node[name=b2]{};
     \draw[rounded corners=6pt] (b1) rectangle (b2);

	\path (3.north west) ++(-0.1,0.1) node[name=d1]{} (4.south east) ++(0.1,-0.1) 
      node[name=d2]{};
     \draw[rounded corners=6pt] (d1) rectangle (d2);
     
      \path (5.north west) ++(-0.1,0.1) node[name=c1]{} (5.south east) ++(0.1,-0.1) 
      node[name=c2]{};
     \draw[rounded corners=6pt] (c1) rectangle (c2);

\path       
      (10,3) node[name=1a]{1}
      (12,5) node[name=2a]{2}
      (12,3) node[name=3a]{3}
      (12,1) node[name=4a]{4}
      (14,5) node[name=5a]{5};
      \path (9,5) node[name=a]{$\mathcal{A}_s$};
      \draw[arrow,shorten >=-5pt, shorten <=-5pt] (1a) to (4a);
      \draw[arrow,shorten >=-3pt, shorten <=-3pt] (2a) to (5a);
                        
     \path (1a.north west) ++(-0.1,0.1) node[name=aa1]{} (1a.south east) ++(0.1,-0.1) 
      node[name=aa2]{};
     \draw[rounded corners=6pt] (aa1) rectangle (aa2);

	\path (2a.north west) ++(-0.1,0.1) node[name=bb1]{} (4a.south east) ++(0.1,-0.1) 
      node[name=bb2]{};
     \draw[rounded corners=6pt] (bb1) rectangle (bb2);
     
      \path (5a.north west) ++(-0.1,0.1) node[name=cc1]{} (5a.south east) ++(0.1,-0.1) 
      node[name=cc2]{};
     \draw[rounded corners=6pt] (cc1) rectangle (cc2);
  \end{tikzpicture}    
\end{center}

\begin{figure}[Ht]
\begin{center}
\normalsize
    \begin{tikzpicture}[scale=0.40]
      \tikzstyle{arrow}=[->,>=latex',shorten >=-3.5pt, shorten <=-3.5pt]
      \path       
      (0,2) node[name=1]{1}
      (2,5) node[name=2]{2}
      (2,3) node[name=3]{3}
      (2,1) node[name=4]{4}
      (2,-1) node[name=5]{5}
      (4,3) node[name=6]{6}
      (4,1) node[name=7]{7}
      (6,3) node[name=8]{8}
      (6,1) node[name=9]{9};
      \path (8,2) node[name=f]{$\Rightarrow$};
      \path (0,5) node[name=a]{$\mathcal{A}$};
      \draw[arrow] (1) to (2);
      \draw[arrow] (1) to (4);
      \draw[arrow] (2) to (6);
      \draw[arrow] (3) to (7);
      \draw[arrow] (4) to (6);
      \draw[arrow] (5) to (7);
      \draw[arrow] (6) to (8);
      \draw[arrow] (7) to (9);
                  
     \path (1.north west) ++(-0.1,0.1) node[name=a1]{} (1.south east) ++(0.1,-0.1) 
      node[name=a2]{};
     \draw[rounded corners=6pt] (a1) rectangle (a2);

	\path (2.north west) ++(-0.1,0.1) node[name=b1]{} (3.south east) ++(0.1,-0.1) 
      node[name=b2]{};
     \draw[rounded corners=6pt] (b1) rectangle (b2);
     
     \path (4.north west) ++(-0.1,0.1) node[name=b3]{} (5.south east) ++(0.1,-0.1) 
      node[name=b4]{};
     \draw[rounded corners=6pt] (b3) rectangle (b4);

     \path (6.north west) ++(-0.1,0.1) node[name=c1]{} (6.south east) ++(0.1,-0.1) 
      node[name=c2]{};
     \draw[rounded corners=6pt] (c1) rectangle (c2);

	\path (7.north west) ++(-0.1,0.1) node[name=c3]{} (7.south east) ++(0.1,-0.1) 
      node[name=c4]{};
     \draw[rounded corners=6pt] (c3) rectangle (c4);

	\path (8.north west) ++(-0.1,0.1) node[name=d1]{} (9.south east) ++(0.1,-0.1) 
      node[name=d2]{};
     \draw[rounded corners=6pt] (d1) rectangle (d2);

  \path       
      (10,2) node[name=1]{1}
      (12,5) node[name=2]{2}
      (12,3) node[name=3]{3}
      (12,1) node[name=4]{4}
      (12,-1) node[name=5]{5}
      (14,3) node[name=6]{6}
      (14,1) node[name=7]{7}
      (16,3) node[name=8]{8}
      (16,1) node[name=9]{9};
      \path (18,2) node[name=f]{$\Rightarrow$};
      \path (10,5) node[name=a]{$\mathcal{A}'$};
      \draw[arrow] (1) to (2);
      \draw[arrow] (1) to (4);
      \draw[arrow] (2) to (6);
      \draw[arrow] (3) to (7);
      \draw[arrow] (4) to (6);
      \draw[arrow] (5) to (7);
      \draw[arrow] (6) to (8);
      \draw[arrow] (7) to (9);

     \path (1.north west) ++(-0.1,0.1) node[name=a1]{} (1.south east) ++(0.1,-0.1) 
      node[name=a2]{};
     \draw[rounded corners=6pt] (a1) rectangle (a2);

	\path (2.north west) ++(-0.1,0.1) node[name=b1]{} (5.south east) ++(0.1,-0.1) 
      node[name=b2]{};
     \draw[rounded corners=6pt] (b1) rectangle (b2);

     \path (6.north west) ++(-0.1,0.1) node[name=c1]{} (6.south east) ++(0.1,-0.1) 
      node[name=c2]{};
     \draw[rounded corners=6pt] (c1) rectangle (c2);

	\path (7.north west) ++(-0.1,0.1) node[name=c3]{} (7.south east) ++(0.1,-0.1) 
      node[name=c4]{};
     \draw[rounded corners=6pt] (c3) rectangle (c4);

	\path (8.north west) ++(-0.1,0.1) node[name=d1]{} (9.south east) ++(0.1,-0.1) 
      node[name=d2]{};
     \draw[rounded corners=6pt] (d1) rectangle (d2);

\path 

      (20,2) node[name=1]{1}
      (22,5) node[name=2]{2}
      (22,3) node[name=3]{3}
      (22,1) node[name=4]{4}
      (22,-1) node[name=5]{5}
      (24,3) node[name=6]{6}
      (24,1) node[name=7]{7}
      (26,3) node[name=8]{8}
      (26,1) node[name=9]{9};
      
      \path (20,5) node[name=a]{$\mathcal{A}''$};
      \draw[arrow] (1) to (2);
      \draw[arrow] (1) to (4);
      \draw[arrow] (2) to (6);
      \draw[arrow] (3) to (7);
      \draw[arrow] (4) to (6);
      \draw[arrow] (5) to (7);
      \draw[arrow] (6) to (8);
      \draw[arrow] (7) to (9);
                               
     \path (1.north west) ++(-0.1,0.1) node[name=a1]{} (1.south east) ++(0.1,-0.1) 
      node[name=a2]{};
     \draw[rounded corners=6pt] (a1) rectangle (a2);

	\path (2.north west) ++(-0.1,0.1) node[name=b1]{} (5.south east) ++(0.1,-0.1) 
      node[name=b2]{};
     \draw[rounded corners=6pt] (b1) rectangle (b2);

     \path (6.north west) ++(-0.1,0.1) node[name=c1]{} (7.south east) ++(0.1,-0.1) 
      node[name=c2]{};
     \draw[rounded corners=6pt] (c1) rectangle (c2);

	\path (8.north west) ++(-0.1,0.1) node[name=d1]{} (9.south east) ++(0.1,-0.1) 
      node[name=d2]{};
     \draw[rounded corners=6pt] (d1) rectangle (d2);

  \end{tikzpicture}    
   \caption{Some abstract transition systems.}
   \label{figure-1}
\end{center}
\end{figure}
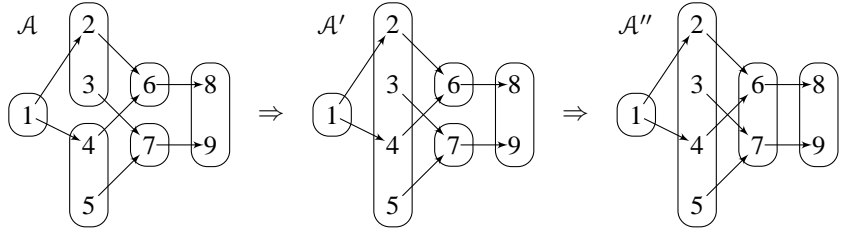

Such a methodology is called EGAS, 
Example-Guided Abstraction Simplification, since this abstraction 
simplification is able to keep examples in the meaning above. 
Let us illustrate how EGAS works through a simple example. 
Let us consider the 
abstract transition system $\cA$ in Figure~\ref{figure-1}, 
where $\{[1],[2,3],[4,5],[6],[7],[8,9]\}$ is the underlying
state partition. The abstract state space of $\cA$ is simplified by merging the
blocks $[2,3]$ and $[4,5]$: EGAS guarantees that 
this can be safely done because 
$\pre^\sharp ([2,3]) =
\pre^\sharp([4,5])= \{[1]\}$
 and $\post^\sharp ([2,3]) =
\post^\sharp([4,5])= \{[6],[7]\}$, where $\pre^\sharp$ and 
$\post^\sharp$ denote, respectively, the abstract predecessor and 
successor functions. 
This abstraction simplification leads to the abstract 
system $\cA'$ in Figure~\ref{figure-1}. 
Observe that the abstract path 
$\pi = \langle [1], [2,3,4,5], [7], [8,9]\rangle$ in $\cA'$
is spurious because there is no concrete path whose 
abstraction in $\cA'$ is  $\pi$, while $\pi$
is instead the abstraction of the spurious path 
$\langle [1],$ $[4,5],$ $[7], [8,9]\rangle$ in $\cA$. 
On the other hand, consider the path
$\sigma = \langle [1], [2,3,4,5], [6], [8,9]\rangle$ in $\cA'$ and
observe that all the paths in $\cA$ 
that are abstracted to $\pi'$, i.e.\
$\langle [1],[2,3],[6],[8,9]\rangle$
and $\langle [1],[4,5],[6],[8,9]\rangle$,  are not spurious. 
This is consistent with the fact that $\sigma$ 
actually is not a spurious path.  
Likewise, $\cA'$ can be further 
simplified to the abstract system $\cA''$ where the blocks 
$[6]$ and $[7]$ are merged. 
This transform  also keeps examples  
because now there is no spurious path in 
$\cA''$. 
Let us also notice that if $\cA$ would get simplified to 
an abstract system $\cA'''$
by merging the blocks $[1]$ and $[2,3]$ 
into a new abstract state $[1,2,3]$ then this transform would
not keep examples because we would obtain the spurious loop path
$\tau = \langle [1,2,3], [1,2,3], [1,2,3], ... \rangle$ in $\cA'''$ 
(because, in $\cA'''$, $[1,2,3]$ would have a self-loop)
while there is no corresponding spurious abstract path  
in $\cA$ whose abstraction in $\cA'''$ is $\tau$. 
 
We show how EGAS can be formalized within the standard 
Galois connection-based 
abstract interpretation framework~\cite{CC77,CC79}.  
Consider for instance the two following 
basic abstract domains $A_1$ and $A_2$ 
for sign analysis of an integer
program variable, so that the concrete domain of values is 
the powerset $\wp(\mathbb{Z})$ of integer numbers.
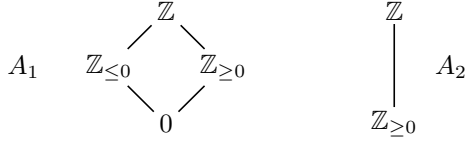
\begin{center}
    \begin{tikzpicture}[scale=0.75,shorten >=-2pt, shorten <=-2pt]
      \draw (0,0) node[name=2] {{$0$}};
      \draw (-1,1) node[name=3] {{$\mathbb{Z}_{\leq 0}$}};
      \draw (1,1) node[name=4] {{$\mathbb{Z}_{\geq 0}$}};
      \draw (0,2) node[name=5] {{$\mathbb{Z}$}};

	\draw (-2.5,1) node {{$A_1$}};

      \draw[semithick] (2) -- (3);
      \draw[semithick] (2) -- (4);
      \draw[semithick] (3) -- (5);
      \draw[semithick] (4) -- (5);
      
      \draw (5,1) node {{$A_2$}};
      
      \draw (4,0) node[name=2] {{$\mathbb{Z}_{\geq 0}$}};
        \draw (4,2) node[name=5] {{$\mathbb{Z}$}};

      \draw[semithick] (2) -- (5);
   
\end{tikzpicture}
\end{center}
Recall that in abstract interpretation 
the best correct approximation 
of a semantic function 
$f$ 
on an abstract domain $A$ 
is given by $f^A \ud \alpha \circ f \circ \gamma$, where 
$\alpha$ and $\gamma$ are
the abstraction and concretization maps defining $A$. 
Let us consider a simple operation of
increment $x$++ on an integer variable $x$. 
In this case, 
the best correct approximations on the abstractions $A_1$ and $A_2$ go
as follows:
\begin{align*}
\text{++}^{A_1} &= \{0 \mapsto \mathbb{Z}_{\geq 0},\: 
\mathbb{Z}_{\leq 0} \mapsto \mathbb{Z},\: 
\mathbb{Z}_{\geq 0} \mapsto \mathbb{Z}_{\geq 0},\: 
\mathbb{Z} \mapsto \mathbb{Z}\},\\[-2.5pt]
\text{++}^{A_2} &= \{
\mathbb{Z}_{\geq 0} \mapsto \mathbb{Z}_{\geq 0},\: 
\mathbb{Z} \mapsto \mathbb{Z}\}.
\end{align*}
We observe that the best correct approximations of 
$\text{++}$ in $A_1$ and $A_2$ encode the same function,
meaning that the approximations of the operation $\text{++}$ in $A_1$ and $A_2$ are
equivalent: In fact, we have that
$\gamma_{A_1} \circ \text{++}^{A_1} \circ \alpha_{A_1}$ and
$\gamma_{A_2} \circ \text{++}^{A_2} \circ \alpha_{A_2}$ are exactly 
the same function in $\wp(\mathbb{Z})\rightarrow \wp(\mathbb{Z})$. 
In other terms, the abstract domain $A_1$ 
contains some ``irrelevant'' abstract values for approximating 
the increment operation, 
namely, $0$ and $\mathbb{Z}_{\leq 0}$. 
We formalize this simplification process of an abstract domain relatively to a 
semantic function in standard Galois connection-based abstract interpretation. 
This allows us to provide, for generic
continuous semantic functions, a systematic and constructive method,
that we call \emph{correctness kernel}, for
simplifying a given abstraction $A$ relatively to a given
semantic function $f$ towards the 
unique minimal abstract domain that
induces an equivalent approximate behavior of $f$ as in $A$.

We show how correctness kernels can be embedded within 
the CEGAR methodology by providing a novel
refinement heuristics in a CEGAR iteration step which turns out to be
more accurate than the basic refinement heuristics~\cite{cgjlv03}.
We also describe how correctness kernels may be applied in 
predicate
abstraction-based model checking \cite{ddp99,gs97} 
for reducing the search space without
applying Ball et al.'s \cite{bpr03} Cartesian abstractions, 
which typically yield additional
loss of precision.

This is an extended and revised version of the conference
paper~\cite{gr10}.

\section{Correctness Kernels}
\subsection{Abstract Interpretation Background}
\paragraph{\textbf{Abstract Domains.}}
In standard abstract interpretation~\cite{CC77,CC79}, 
abstract domains (or abstractions) 
are specified by Galois connections/insertions
(GCs/GIs for short) or, equivalently, adjunctions. 
Concrete and abstract domains, $\tuple{C,\leq_C}$ and
$\tuple{A,\leq_A}$, are assumed to be complete lattices 
which are related by abstraction and concretization maps
$\alpha:C\ra A$ and $\gamma:A \ra C$ that give
rise to an adjunction $(\alpha,C,A,\gamma)$, that is,
for all $a$ and $c$,
$\alpha(c) \leq_A a \Lra c \leq_C \gamma(a)$. A GC is a GI when $\alpha\comp\gamma=\lambda x.x$.
It is known that the function
$\ok{\mu_A \ud \gamma \circ \alpha: C \ra C}$ is an upper closure operator (uco) on $C$, i.e.\ a
monotone, idempotent and increasing function. 
GIs of a common concrete domain $C$ are preordered w.r.t.\ their relative precision as usual: 
$\cG_1 = (\alpha_1,C,A_1,\gamma_1)\sqsubseteq \cG_2=(\alpha_2,C,A_2,\gamma_2)$~---~i.e.\ $A_1$/$A_2$ is a refinement/simplification of  $A_2$/$A_1$~---~iff
$\gamma_2 (\alpha_2(C)) \subseteq\gamma_1 (\alpha_1 (C))$.
Moreover, $\cG_1$ and $\cG_2$ are equivalent when $\cG_1
\sqsubseteq \cG_2$ and $\cG_2 \sqsubseteq \cG_1$. 
We denote by $\Abs(C)$ the family of abstract domains of $C$ up to the above
equivalence. 
It is well known that
$\tuple{\Abs(C),\sqsubseteq}$ is a complete lattice.  
Thus, one can consider the most concrete simplification (i.e., lub $\sqcup$) and 
the most abstract refinement (i.e., glb $\sqcap$) 
of any family of abstract domains.
Abstract domains can be
equivalently defined as uco's, meaning that any 
GI $(\alpha,C,A,\gamma)$ induces the uco $\mu_A$, any uco $\mu: C\ra C$ induces the GI $(\mu, C, \mu(C), \lambda x.x)$, and these two 
transforms are the inverse of each other, namely $(\mu_A,C,\mu_A(C), \lambda x.x)$ and
$(\alpha,C,A,\gamma)$ are equivalent GIs and $\mu = \mu_{\mu(C)}$. 
In more technical terms,  $\tuple{\Abs(C),\sqsubseteq}$
is isomorphic to the complete lattice $\tuple{\uco(C),\sqsubseteq}$ of uco's on $C$, 
where $\sqsubseteq$
denotes the standard point-wise ordering between functions, 
so that lub's and glb's of 
abstractions can be equivalently characterized in $\uco(C)$. 
Let us also recall that each closure $\mu\in \uco(C)$ is
uniquely determined by its image $\img(\mu)=\mu(C)$ as follows: for any $x\in C$,
$\mu(x) = \wedge \!\{ y\in C~|~ y\in \mu(C), \, x\leq y\}$.  
Moreover, a
subset $X\subseteq C$ is the image of some uco on $C$ iff $X$
is meet-closed, i.e.\ $X=\ok{\Cl_\wedge 
(X)}\ud \{ \wedge Y~|~ Y\subseteq
X\}$ (note~that $\top_C =\wedge \varnothing \in \Cl_\wedge (X)$).
This allows us to equivalently use uco's both as functions in $C\ra C$ or as subsets 
of $C$; this
does not give rise to ambiguity, since one can distinguish their use
as functions or sets according to the context. 
Hence, if $A,B\in \Abs(C)$ are two abstractions of $C$ then they 
can be viewed as images of two uco's on $C$, denoted
respectively by $\mu_A$ and $\mu_B$, so that 
$A$ is a refinement of
$B$ when $\img(\mu_B) \subseteq \img(\mu_A)$. Let us also recall
that given a family of uco's $\mathcal{X}\subseteq \uco(C)$, then its lub and glb can be characterized 
as follows: $\sqcup\, \mathcal{X} = \cap_{\mu\in \mathcal{X}} \img(\mu)$ and 
$\sqcap\, \mathcal{X} = \Cl_\wedge\big( \cup_{\mu\in \mathcal{X}} \img(\mu)\big)$.

\paragraph{\textbf{Abstract Functions.}}
Let $f:C\ra C$ be some 
concrete semantic function~---~for simplicity, 
we consider 1-ary functions~---~and let
 $\ok{f^\sharp:A \ra A}$ be a corresponding abstract function
 defined on some abstraction $A\in \Abs(C)$. Then,
$\ok{\tuple{A,f^\sharp}}$ is a sound abstract interpretation
when $\ok{\alpha \circ f \sqsubseteq f^\sharp\circ \alpha}$ holds.
Moreover, the abstract function
$\ok{f^A \ud \alpha \circ f \circ \gamma: A\rightarrow A}$ is called the \emph{best
correct approximation} (b.c.a.) of $f$ on $A$ because any abstract
interpretation
$\ok{\tuple{A,f^\sharp}}$ is sound iff for any $a\in A$, 
$\ok{f^A(a) \leq f^\sharp(a)}$. Hence, for any abstraction $A$, 
$\ok{f^A}$ plays the role of the 
best possible approximation of $f$ on the abstract domain $A$. 

\subsection{The Problem}
Given a semantic function $f:C\rightarrow C$ on some concrete domain 
$C$ and 
an abstraction 
$A \in \Abs(C)$, does there exist the \emph{most abstract domain} that induces the same 
best correct approximation of $f$ as $A$ does? 

Let us formalize the above question. 
Consider two abstractions $A,B\in \Abs(C)$. We say that
$A$ and $B$ induce the same best correct approximation
of $f$ when  $f^A$ and $f^B$ approximate
any concrete computation $f(c)$ in the same way, namely, for any $c\in C$, 
$\gamma_A (f^A (\alpha_A (c))) = \gamma_B (f^B (\alpha_B (c)))$. 
By recalling that $\mu_A$ and $\mu_B$ denote the corresponding uco's, this definition 
boils down to the following equation:
$$\mu_A \comp f \comp \mu_A = \mu_B \comp f \comp \mu_B .$$
In order to keep the notation easy, this is denoted simply by $\ok{f^A = f^B}$. 
Also, if $F\subseteq C\ra C$ is a set of concrete functions then 
$\ok{F^A = F^B}$ means that for any $f\in F$, $\ok{f^A = f^B}$. 

Given $A\in \Abs(C)$, the domain  
$\ok{\sqcup \{B \in \Abs(C)~|~ F^B = F^A\}}$
is precisely the lub in $\Abs(C)$ of all the abstractions that induce the same
best correct approximations of $F$ as $A$ does.   
Hence, our question is formalized 
through the following notion of correctness kernel.

\begin{definition}
\rm
Given $F\subseteq C\rarr{} C$, define
$\mathscr{K}_F: \Abs(C) \rightarrow \Abs(C)$ as 
$$\mathscr{K}_F(A) \triangleq \sqcup \{ B \in \Abs(C) ~|~ F^B = F^A\}.$$
\noindent
If $F^{\mathscr{K}_F(A)} = F^A$ 
then $\mathscr{K}_F(A)$ is called the \emph{correctness kernel} of $A$ for $F$. \qed
\end{definition}

A correctness kernel, when it exists, is an abstraction simplification. 
It is worth observing that 
the corresponding dual abstraction refinement does not exist, 
namely, the dual question on the 
existence of the most concrete abstraction that induces 
the same best correct approximation of $f$ as $A$ has a negative answer, 
as shown by the following simple example. 

\begin{example}\label{esempio2}
\rm
Consider the lattice $C$ depicted below.
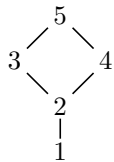
\begin{center}
    \begin{tikzpicture}[scale=0.6,shorten >=-2pt, shorten <=-2pt]
      \draw (0,0) node[name=1] {$1$};
      \draw (0,1) node[name=2] {$2$};
      \draw (-1,2) node[name=3] {$3$};
      \draw (1,2) node[name=4] {$4$};
      \draw (0,3) node[name=5] {$5$};

      \draw[semithick] (1) -- (2);
      \draw[semithick] (2) -- (3);
      \draw[semithick] (2) -- (4);
      \draw[semithick] (3) -- (5);
      \draw[semithick] (4) -- (5);
\end{tikzpicture}
\end{center}
Let us consider 
the monotonic function $f:C\rightarrow C$ defined as $f\triangleq \{1\mapsto 1,\, 2\mapsto 1,\, 3\mapsto 5,\,$ $4\mapsto 5,\, 5\mapsto 5\}$. Let us consider the abstraction $A\in \Abs(C)$ 
defined as $A=\{1,5\}$, so that the corresponding uco 
$\mu \in \uco(C)$ is the  function: 
$\mu = \{1\mapsto 1,\, 2\mapsto 5,\, 3\mapsto 5,\,$ $4\mapsto 5,\, 5\mapsto 5\}$.
It is immediate to  observe 
that $\mu\comp f \comp \mu = \{1\mapsto 1,\, 2\mapsto 5,\, 3\mapsto 5,\, 4\mapsto 5,\, 5\mapsto 5\}$.
Consider now the abstractions 
$\rho_1 \triangleq \{1,3,5\}$  and $\rho_2 \triangleq \{1,4,5\}$ and observe that $\rho_i \comp f \comp \rho_i = \mu\comp f \comp \mu$. However, we have that 
$\rho_1 \sqcap \rho_2 = \lambda x.x$, because the image
of $\rho_1 \sqcap \rho_2$ is $\Cl_\wedge (\rho_1 \cup \rho_2) = 
\{1,2,3,4,5\}$. Hence, $(\rho_1 \sqcap \rho_2) \comp f \comp 
 (\rho_1 \sqcap \rho_2) = f \neq \mu\comp f \comp \mu$. Let $\rho_r \
 ud \sqcap \{ \rho \in \uco(C)~|~ \rho \circ f \circ 
 \rho = \mu \circ f \circ \mu\}$.
 Thus, $\rho_r = \lambda x.x$ and, in turn,   
 $\rho_r \comp f \comp \rho_r
 \neq \mu\comp f \comp \mu$.
 Consequently, the 
most concrete abstraction
that induces the same best correct approximation of $f$ as $\mu$ does not exist. \qed
\end{example}

\section{Characterization of Correctness Kernels}

Our key technical result provides a \emph{constructive} characterization 
of the property of ``having the same b.c.a.'' for two comparable abstract domains.
In the following, given a poset $A$ and any subset 
$S\subseteq A$, $\max (S) \triangleq \{
x\in S~|~ \forall y\in S.\; x\leq_A y \Rightarrow x = y\}$ denotes the
set of maximal elements of $S$ in $A$.

\begin{lemma}\label{key}
Let $f: C\rightarrow C$ and  
$A,B \in \Abs(C)$ such that $A\sqsubseteq B$. Assume that the function
$f\comp \mu_A : C \ra C$ is continuous (i.e., preserves lub's of chains in $C$). Then, 
$$\textstyle
f^B = f^A \;\Lra\; \gamma_A\Big(\img(f^A) \cup \bigcup_{y\in A} 
\max(\{x\in A~|~f^A(x) \leq_A y\})\Big) 
\subseteq \gamma_B\big(B\big).$$
\end{lemma}
\begin{proof}
Let $\mu$ and $\rho$ be the uco's on $C$ induced by, respectively, the abstractions
$A$ and $B$, so that $\mu \sqsubseteq \rho$. Let us recall (see e.g.\ \cite[Proposition~4.2.3.0.1]{cou78})
that $\mu \sqsubseteq \rho$ implies
$\mu \circ \rho = \rho = \rho \circ \mu$. 

\noindent
Given $y\in A$, let us show that
$$\gamma_A\big(\max(\{x\in A ~|~ f^A(x) \leq_A y\})\big) = 
\max(\{x\in C ~|~ f  (\mu (x)) \leq_C \gamma_A(y)\}).$$ 
\noindent
$(\subseteq)$  Let
$z\in \max(\{x\in A ~|~ f^A (x) \leq_A y\})$. 
Then, $\alpha_A (f(\gamma_A (z))) \leq_A y$ iff
$f(\gamma_A (z)) \leq_C \gamma_A(y)$ iff $f(\gamma_A(\alpha_A (\gamma_A (z)))) \leq_C \gamma_A (y)$ so that
$\gamma_A(z)\in 
\{x\in C ~|~ f  (\mu (x)) \leq_C \gamma_A(y)\}$. 
Consider $w\in \{x\in C ~|~ f  (\mu (x)) \leq  \gamma_A(y)\}$ such that  
$\gamma_A(z)\leq w$. Thus, 
since $f(\mu(w))\leq_C \gamma_A(y)$ iff $f^A(\alpha_A (w)) \leq_A y$ and
and $z\leq_A \alpha_A(w)$, by maximality 
of $z$, $z=\alpha_A(w)$ so that $w\leq \gamma_A(\alpha_A(w)) = 
\gamma_A(z)$ and in turn $\gamma_A(z)=w$. Therefore, 
$\gamma_A(z)\in \max(\{x\in C ~|~ f  (\mu (x)) \leq_C \gamma_A(y)\})$.  

\noindent
$(\supseteq)$
Let $z\in \max(\{x\in C ~|~ f  (\mu (x)) \leq_C \gamma_A(y)\})$. Then, 
$f(\gamma_A(\alpha_A(z))) \leq_C \gamma_A (y)$, hence we have that 
$\alpha_A(f(\gamma_A (\alpha_A (z)))) \leq_A y$, and in turn
 $\alpha_A(z) \in \{x\in A ~|~ f^A(x) \leq_A y\}$. 
Consider $w\in \{x\in A ~|~ f^A(x) \leq_A y\}$ such that $\alpha_A (z) \leq_A w$. Then, 
$\gamma_A (w) \in \{x\in C ~|~ f  (\mu (x)) \leq_C \gamma_A(y)\}$, so that from $z\leq_C \gamma_A (w)$,
by maximality 
of $z$, we obtain
$z = \gamma_A(w)$. Thus, $\alpha_A (z)=\alpha_A(\gamma_A(w)) = w$, so that 
$\alpha_A(z) \in \max(\{x\in A ~|~ f^A(x) \leq_A y\})$. Moreover, 
$f(\gamma_A (\alpha_A (\gamma_A (\alpha_A (z))))) = f((\gamma_A (\alpha_A (z)))) \leq_C  \gamma_A (y)$
so that
$\gamma_A(\alpha_A (z)) \in \{x\in C ~|~ f  (\mu (x)) \leq_C \gamma_A(y)\}$. Thus,  
since $z\leq_C \gamma_A(\alpha_A (z))$ 
by maximality
of $z$, $z=\gamma_A(\alpha_A (z))$. Therefore, 
$z\in \gamma_A\big(\max(\{x\in A ~|~ f^A(x) \leq_A y\})\big)$.

\medskip
\noindent
Thus, if $\downarrow\! y \ud \{x\in C~|~x\leq y\}$, then
$ \gamma_A\big(\max(\{x\in A ~|~ f^A(x) \leq_A y\})\big) =
\max ((f \comp \mu)^{-1}(\ok{\downarrow \!\gamma_A(y)}))$. Also, 
note that $\gamma_A(\img(f^A)) = \gamma_A(\alpha_A(f(\gamma_A(A)))) = 
\gamma_A(\alpha_A(f(\gamma_A(\alpha_A(C)))))=
\mu(f(\mu(C)))$.
We therefore prove the following equivalent 
statement which is formalized through uco's: 
$$\rho \comp f \comp \rho = \mu \comp f \comp \mu
\text{~~iff~~}
\mu(f(\mu(C))) \cup 
\textstyle \bigcup_{y\in \mu} \max((f \comp \mu)^{-1}(\downarrow\! y))\subseteq \rho.$$
Let us first prove that $$\rho \comp f \comp \rho = \mu \comp f \comp \mu \:\Leftrightarrow\: \rho \comp f \comp \mu = \mu \comp f \comp \mu = \mu \comp f \comp \rho\eqno(*)$$ 

\noindent
($\Rightarrow$) 
On the one hand,
\begin{align*}
\mu \comp f \comp \mu = \rho \comp f \comp \rho  & \Rightarrow 
\text{~~~~\big[by applying $\rho$ to both sides\big]}
\\
\rho \comp \mu \comp f \comp \mu = \rho\comp \rho \comp f \comp \rho 
& \Rightarrow \text{~~~~\big[because $\rho \comp \mu = \rho$ and 
$\rho \comp \rho = \rho$\big]} \\
\rho  \comp f \comp \mu = \rho \comp f \comp \rho &\Rightarrow 
\text{~~~~\big[by hypothesis\big]}
\\
\rho  \comp f \comp \mu = \mu \comp f \comp \mu &
\end{align*}
and on the other hand,
\begin{align*}
\mu \comp f \comp \mu = \rho \comp f \comp \rho  & \Rightarrow 
\text{~~~~\big[by applying $\rho$ in front to both sides\big]}
\\
\mu \comp f \comp \mu \comp \rho = \rho\comp f \comp \rho \comp \rho 
& \Rightarrow 
\text{~~~~\big[because $\mu \comp \rho = \rho$ and $\rho \comp \rho = \rho$\big]} \\
\mu  \comp f \comp \rho = \rho \comp f \comp \rho &\Rightarrow 
\text{~~~~\big[by hypothesis\big]}
\\
\mu  \comp f \comp \rho = \mu \comp f \comp \mu &
\end{align*}
so that $\rho \comp f \comp \mu = \mu \comp f \comp \mu = \mu \comp f \comp \rho$.  

\medskip
\noindent
($\Leftarrow$) We have that:
\begin{align*}
\rho \comp f \comp \mu = \mu \comp f \comp \rho & \Rightarrow 
\text{~~~~\big[by applying $\rho$ to both sides\big]}
\\
\rho \comp \rho \comp f \comp \mu = \rho \comp \mu \comp f \comp \rho & \Rightarrow \text{~~~~\big[since $\rho \comp \rho = \rho$ and 
$\rho \comp \mu = \rho$\big]}\\
\rho \comp f \comp \mu = \rho \comp f \comp \rho & \Rightarrow
\text{~~~~\big[by hypothesis\big]}
\\
\mu \comp f \comp \mu = \rho \comp f \comp \rho. &
\end{align*}

\noindent
Let us now observe that $\rho \comp f \comp \mu = \mu \comp f \comp \mu
\Leftrightarrow \mu (f(\mu(C))) \subseteq \rho$: 
in fact, since $\rho = \rho \comp \mu$, we have that 
$\rho \comp f \comp \mu = \mu \comp f \comp \mu
\Leftrightarrow 
\rho \comp \mu \comp f \comp \mu = \mu \comp f \comp \mu$, and this latter equation
is clearly equivalent to $\mu (f(\mu(C))) \subseteq \rho$. 

\noindent
Moreover, 
since $\rho = \mu \comp \rho$, we have that $\mu \comp f \comp \mu = \mu \comp f \comp \rho$ is equivalent to $\mu \comp (f \comp \mu) = \mu \comp (f\comp \mu) \comp \rho$. This
latter equation states the completeness of the pair of abstractions 
$\langle\rho,\mu\rangle$ for the function $f\comp \mu$. 
By the characterization 
of completeness in \cite[Lemma~4.2]{grs00}, 
since, by hypothesis,  
$f \comp \mu$ is continuous, we have that 
the completeness equation 
$\mu \comp (f \comp \mu) = \mu \comp (f\comp \mu) \comp \rho$
 is equivalent to 
$\cup_{y\in \mu} \max((f \comp \mu)^{-1}(\downarrow\! y))\subseteq \rho$. 
Thus, $\mu \comp f \comp \mu = \mu \comp f \comp \rho \Leftrightarrow 
\cup_{y\in \mu} \max((f \comp \mu)^{-1}(\downarrow\! y))\subseteq \rho$. 

\noindent
Summing up, we have shown that 
$$\rho \comp f \comp \mu = \mu \comp f \comp \mu = \mu \comp f \comp \rho \:\Leftrightarrow\: \mu (f(\mu(C))) \cup \textstyle 
\bigcup_{y\in \mu} \max((f \comp \mu)^{-1}(\downarrow\! y))\subseteq \rho$$
By the above equivalence~$(*)$, this implies the thesis.
\end{proof}

\begin{remark}
\rm
It is important to stress that the above proof of Lemma~\ref{key}
basically consists in reducing the equality $f^A = f^B$ between b.c.a.'s to a
standard 
property of completeness of the abstract domains $A$ and $B$ for the function $f$ and
then in exploiting the constructive characterization of completeness
of abstract domains given by Giacobazzi et al.~\cite[Section~4]{grs00}.
In this sense, this proof provides an unexpected reduction of an equivalence problem 
between best correct approximations 
to a completeness problem.
This is particularly interesting because while best
correct approximations can be always defined on any abstract domain,
complete approximations are instead uncommon since they represent an ideal
situation where fixed point computations of complete approximations do
not loose precision \cite{CC77,grs00}. \qed
\end{remark}

As a consequence of Lemma~\ref{key} we obtain the following constructive result of existence for correctness kernels. 

\begin{theorem}\label{kernel}
Let $A \in \Abs(C)$ and $F\subseteq C\rightarrow C$ such that, for
any $f\in F$, 
$f \comp \mu_A$ is continuous. Then, the
correctness kernel of $A$ for $F$ exists and it is
$$
\mathscr{K}_F(A) = \Clv\Big(\bigcup_{f\in F} \big(\textstyle \img(f^A) \cup  \bigcup_{y\in \img(f^A)} 
\max(\{x\in A~|~f^A(x) = y\})\big)\Big).$$
\end{theorem}
\begin{proof}
Let $\mu = \mu_A$. Let us first prove that the correctness kernel of $A$ for $F$ exists,
namely $F^{\mathscr{K}_F(\mu)} = F^\mu$. Since $\mu
\sqsubseteq \mathscr{K}_F(\mu)$, by Lemma~\ref{key}, it is sufficient to show that 
for any $f\in F$, 
$$\mu (f (\mu (C))) \cup {\textstyle \bigcup_{y\in \mu}} \max(\{x\in C ~|~
f (\mu(x)) \leq y\}) \subseteq \mathscr{K}_F(\mu) = \cap \{\rho \in \uco(C)~|~  \rho f\rho  = \mu f\mu\}.$$ 
We therefore consider $\rho \in \uco(C)$ such that $\rho f \rho = \mu f \mu$ and we prove that 
$\mu (f (\mu (C))) \cup \bigcup_{y\in \mu} \max(\{x\in \mu ~|~
f (\mu( x)) \leq y\}) \subseteq \rho(C)$. From $\mu f \mu=\rho f \rho $ by applying $\rho$ we obtain
$\rho \mu f \mu = \rho \rho f \rho = \rho f \rho = \mu f \mu$, so that 
$\mu (f (\mu(C)))\subseteq \rho(C)$. Moreover, from $\mu f \mu=\rho f \rho $ by applying 
$\rho$ in front, we obtain $\mu f \mu \rho = \rho f \rho \rho = \rho f \rho = \mu f \mu$. 
As done in the proof of Lemma~\ref{key}, 
by the characterization 
of completeness in \cite[Lemma~4.2]{grs00}, 
since, by hypothesis,  
$f \mu$ is continuous, we have that 
$\mu (f \mu) \rho = \mu (f \mu)$ implies
$\bigcup_{y\in \mu} \max(\{x\in C ~|~
f (\mu( x)) \leq y\}) \subseteq \rho$. 

\noindent
Hence,  the correctness kernel of $A$ for $F$ exists.
Next, we prove that 
$$
\mathscr{K}_F(\mu) = 
\mathscr{K}'_F(\mu) \ud
\sqcup \{\rho \in \uco(C)~|~ \rho \sqsupseteq \mu,\: F^\rho = F^\mu\}.\eqno(*)
$$
In fact, since $\{\rho \in \uco(C)~|~ \rho \sqsupseteq \mu,\: F^\rho = F^\mu\}
\subseteq \{\rho \in \uco(C)~|~  F^\rho = F^\mu\}$, we have that 
$\mathscr{K}'_F(\mu) \sqsubseteq \mathscr{K}_F(\mu)$. On the other hand, since 
$F^{\mathscr{K}_F(\mu)} = F^\mu$ and  
$\mathscr{K}_F(\mu) \sqsupseteq \mu$, we also have that 
 $\mathscr{K}_F(\mu) \in \{\rho \in \uco(C)~|~ \rho \sqsupseteq \mu,\: F^\rho = F^\mu\}$
and therefore $\mathscr{K}_F(\mu) \sqsubseteq \mathscr{K}'_F(\mu)$. 

\noindent
We now consider the following chain of equalities:
\begin{align*}
\mathscr{K}_F(\mu) =\\[3pt] 
\text{\big[by equation $(*)$\big]~~~~~}\\[3pt]
{\textstyle\bigsqcup} \{\rho \in \uco(C)~|~ \rho \sqsupseteq \mu,\: 
 F^\rho = F^\mu \} = \\[3pt]
\text{\big[by a characterization of lub of uco's\big]~~~~~}\\[3pt]
{\textstyle\bigcap} \{\rho \in \uco(C)~|~ \rho \subseteq \mu,\:  
 F^\rho = F^\mu\} = \\[3pt]
\text{\big[by Lemma~\ref{key}\big]~~~~~}\\[3pt]
{\textstyle\bigcap} \{\rho \in \uco(C)~|~ \rho \subseteq \mu,\:
 {\textstyle \bigcup_{f\in F}} \big( \mu (f(\mu(C))) \cup \textstyle 
\bigcup_{y\in \mu} \max((f \mu)^{-1}(\downarrow\! y))\big) \subseteq \rho
\} =\\[3pt]
\text{\big[because  $\textstyle \bigcup_{f\in F} \big(\mu (f(\mu(C))) \cup \textstyle 
\bigcup_{y\in \mu} \max((f \mu)^{-1}(\downarrow\! y))\big) \subseteq$~~~~~~}\\
\text{$\textstyle \Cl_\wedge \Big(\bigcup_{f\in F} \big(\mu (f(\mu(C))) \cup \textstyle 
\bigcup_{y\in \mu} \max((f \mu)^{-1}(\downarrow\! y))\big)\Big)\subseteq \mu$\big]~~~~~}\\[3pt]
\textstyle \Cl_\wedge \Big(\bigcup_{f\in F} \big(\mu (f(\mu(C))) \cup \textstyle 
\bigcup_{y\in \mu} \max((f \mu)^{-1}(\downarrow\! y))\big)\Big). \phantom{=}
\end{align*}
To conclude, let us show that 
\begin{multline*}
\textstyle \Cl_\wedge \Big(\bigcup_{f\in F} \big(\mu (f(\mu(C))) \cup \textstyle 
\bigcup_{y\in \mu} \max((f \mu)^{-1}(\downarrow\! y))\big)\Big)
=\\
\textstyle \gamma \Big( \Clv\Big(\bigcup_{f\in F} \big(\textstyle \img(f^A) \cup  \bigcup_{y\in \img(f^A)} 
\max(\{x\in A~|~f^A(x) = y\})\big)\Big) \Big).
\end{multline*}
Since $\gamma$ preserves arbitrary glb's (see e.g.\ \cite[Theorem~4.2.7.0.3]{cou78}), it is enough
to show that for any $f\in F$, 
\begin{multline*}
\textstyle \mu (f(\mu(C))) \cup \textstyle 
\bigcup_{y\in \mu} \max((f \mu)^{-1}(\downarrow\! y))
=
\textstyle \gamma \big( \img(f^A) \cup  \bigcup_{y\in \img(f^A)} 
\max(\{a\in A~|~f^A(a) = y\})\big).
\end{multline*}
Firstly, since $\alpha(C)=A$ (see e.g.\ \cite[Theorem~4.2.7.0.3]{cou78}),
we have that $\gamma (\img (f^A)) = \gamma (\alpha (f(\gamma (A)))) =
\gamma (\alpha (f(\gamma(\alpha(C))))) = \mu(f(\mu(C)))$.
Next, we show that 
$$
\textstyle 
\bigcup_{y\in \mu} \max((f \mu)^{-1}(\downarrow\! y))
=\gamma \big(\bigcup_{y\in \img(f^A)} 
\max(\{a\in A~|~f^A(a) = y\})\big)
$$

\noindent
$(\subseteq)$: Consider $y\in \mu$ and $z\in  \max(\{x\in C~|~ f (\mu (x)) \leq y\})$.
Let us first observe that $z=\mu(z)$: in fact, since $f(\mu(\mu(z))) = f(\mu(z)) \leq y$ 
and $z\leq \mu(z)$, by maximality of $z$, $z=\mu(z)$. 
Then, define $y_z \ud f^A(\alpha(z)) \in \img(f^A)$ and let us  show that 
$\alpha (z) \in \max(\{a\in A~|~f^A(a) = y_z\})$. First, $\alpha(z) \in 
\{a\in A~|~f^A(a) = y_z\}$ by definition. Then, consider some $b\in A$ such that 
$f^A(b)=y_z$ and $\alpha(z)\leq b$. We have that $z\leq \gamma(b)$ and 
$\alpha(f(\gamma(b))) = \alpha(f(\gamma(\alpha(z)))) \leq \alpha(y)$, so that 
$f(\mu(\gamma(b)))=f(\gamma(b)) \leq \gamma(\alpha(y))=y$. Hence, by maximality of $z$, 
from $z\leq \gamma(b)$ we obtain $z=\gamma(b)$, and in turn $\alpha(z)=\alpha(\gamma(b))=b$. 
Hence, from $\alpha (z) \in \max(\{a\in A~|~f^A(a) = y_z\})$, since $\gamma(\alpha(z))=z$, we derive
that $z\in \gamma(\max(\{a\in A~|~f^A(a) = y_z\}))$.

\noindent
$(\supseteq)$: Consider $y=f^A(b)$ for some $b\in A$, 
and $z\in \gamma(\max(\{a\in A~|~f^A(a) = y\}))$. Therefore, $z=\gamma(a)$, for some $a\in \max(\{a\in A~|~f^A(a) = y\})$. 
Because $\gamma(y)\in \mu$, 
let us show that $z\in 
\ok{\max((f \mu)^{-1}(\downarrow\! \gamma(y)))}$.
Since $\alpha(f(\gamma(a)))=y$, we have that $f(\mu(z))=f(\gamma(\alpha(z)))=f(\gamma(a)) \leq \gamma(y)$, 
and in turn we derive $z\in  \ok{(f \mu)^{-1}(\downarrow\! \gamma(y))}$. If $f(\mu(u)) \leq \gamma(y)$ and $z\leq u$
then we have that 
$y=f^A(\alpha(z)) \leq f^A(\alpha(u)) = 
\alpha(f(\gamma(\alpha(u)))) = \alpha(f(\mu(u))) \leq \alpha(\gamma(y))=y$, so that
$f^A(\alpha(u))=y$. Hence, from $z=\gamma(a)\leq u$, we obtain $a\leq \alpha(u)$, so that,
by maximality of $a$, $a=\alpha(u)$, namely, $\alpha(z)=\alpha(u)$. Hence, $z=\gamma(\alpha(z))=
\gamma(\alpha(u))$. Hence, $u\leq \gamma(\alpha(u))=z$, from which $z=u$. We can thus conclude that
$z\in \max((f \mu)^{-1}(\downarrow\! \gamma(y)))$.
\end{proof}

Let us illustrate through a simple numerical example how to use the above result
for deriving correctness kernels.  

\begin{example}\rm 
Consider sets of integers 
$\tuple{\wp(\mathbb{Z}),\subseteq}$ as concrete 
domain and a collecting square 
operation $\sq:\wp(\mathbb{Z}) \rightarrow \wp(\mathbb{Z})$ 
as concrete function, i.e., $\sq(X) \triangleq \{x^2 ~|~ x\in X\}$, which is obviously additive and therefore continuous. 
Consider the abstract domain $\Sign \in \Abs(\wp(\mathbb{Z}))$,
depicted in the following diagram, 
that represents the sign of an integer 
variable. 
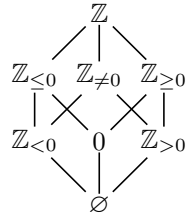
\begin{center}
{
    \begin{tikzpicture}[scale=0.85,shorten >=-2.5pt, shorten <=-2.5pt]
      \draw (0,0) node[name=1] {{$\varnothing$}};

      \draw (-1,1) node[name=2] {{$\mathbb{Z}_{<0}$}};
      \draw (0,1) node[name=3] {{$0$}};
      \draw (1,1) node[name=4] {{$\mathbb{Z}_{>0}$}};

      \draw (-1,2) node[name=5] {{$\mathbb{Z}_{\leq 0}$}};
      \draw (0,2) node[name=6] {{$\mathbb{Z}_{\neq 0}$}};
      \draw (1,2) node[name=7] {{$\mathbb{Z}_{\geq 0}$}};

      \draw (0,3) node[name=8] {{$\mathbb{Z}$}};

      \draw[semithick] (1) -- (2);
      \draw[semithick] (1) -- (3);
      \draw[semithick] (1) -- (4);
      \draw[semithick] (2) -- (5);
      \draw[semithick] (2) -- (6);
      \draw[semithick] (3) -- (5);
      \draw[semithick] (3) -- (7);
      \draw[semithick] (4) -- (6);
      \draw[semithick] (4) -- (7);
      \draw[semithick] (5) -- (8);
      \draw[semithick] (6) -- (8);
      \draw[semithick] (7) -- (8);
\end{tikzpicture}
}
\end{center}
It is immediate to check that $\Sign$ induces the following best correct
approximation of $\sq$:  
\begin{align*}
\sq^{\Sign} = \{&\varnothing \mapsto \varnothing, \mathbb{Z}_{<0} \mapsto 
\mathbb{Z}_{> 0}, 0 \mapsto 0, \mathbb{Z}_{> 0} \mapsto
\mathbb{Z}_{> 0}, \mathbb{Z}_{\leq 0} \mapsto \mathbb{Z}_{\geq 0}, \\
& \mathbb{Z}_{\neq 0} \mapsto \mathbb{Z}_{>0},
\mathbb{Z}_{\geq 0} \mapsto \mathbb{Z}_{\geq 0},
\mathbb{Z} \mapsto \mathbb{Z}_{\geq 0}\}.
\end{align*}
Let us characterize the correctness kernel 
$\mathscr{K}_{\sq}(\Sign)$ by Theorem~\ref{kernel}.
We have that $\img(\sq^{\Sign})= \{\varnothing, \mathbb{Z}_{>0}, 0, 
\mathbb{Z}_{\geq 0}\}$. Moreover, 
\begin{align*}
\max (\{x\in \Sign~|~ \sq^{\Sign}(x) = \varnothing\}) &= \{\varnothing\}\\
\max (\{x\in \Sign~|~ \sq^{\Sign}(x) = \mathbb{Z}_{>0}\}) &= \{\mathbb{Z}_{\neq 0}\}\\ 
\max (\{x\in \Sign~|~ \sq^{\Sign}(x) = 0\}) &= \{0\}\\
\max (\{x\in \Sign~|~ \sq^{\Sign}(x) = \mathbb{Z}_{\geq 0}\}) &= \{\mathbb{Z}\}
\end{align*} 
Therefore, $\bigcup_{y\in \img(\sq^{\Sign})} \max (\{x\in \Sign~|~ \sq^{\Sign}(x) = y\}) 
= \{\varnothing,\mathbb{Z}_{\neq 0},0,\mathbb{Z}\}$ so that,
by Theorem~\ref{kernel}: 
$$\mathscr{K}_{\sq}(\Sign) = \Cl_{\cap}(\{\varnothing, \mathbb{Z}_{> 0}, 
0,\mathbb{Z}_{\geq 0}, \mathbb{Z}_{\neq 0}, \mathbb{Z}\}) = \Sign \smallsetminus
\{\mathbb{Z}_{< 0}, \mathbb{Z}_{\leq 0}\}.$$
Thus, it turns out that we can safely remove the abstract values 
$\mathbb{Z}_{< 0}$ and $\mathbb{Z}_{\leq 0}$
from $\Sign$ and still preserve the same b.c.a.\ as $\Sign$. Besides, we cannot remove further abstract elements otherwise we do not retain the same b.c.a.\ as $\Sign$. For example, this means that $\Sign$-based analyses of programs like 
\[
x := k;
\textbf{while}~\text{condition}~ \textbf{do}
~x := x*x;
\]
can be carried out by using 
the simpler domain $\Sign \smallsetminus
\{\mathbb{Z}_{< 0}, \mathbb{Z}_{\leq 0}\}$, yet 
providing the same input/output abstract behavior. \qed
\end{example}

It is  worth remarking that in Theorem~\ref{kernel} the hypothesis 
of continuity is crucial for the existence of correctness kernels
as the following example shows.

\begin{example}\label{esempio1}\rm
Let us consider the concrete domain $C$ depicted below, namely the ordinal numbers 
less than or equal to $\omega+1$. 
\begin{center}
{
    \begin{tikzpicture}[scale=0.55,shorten >=-1.5pt, shorten <=-1.5pt]
      {\small \draw (0,0) node[name=0] {{$0$}};
      \draw (0,1) node[name=1] {{$1$}};
      \draw (0,2) node[name=2] {{$2$}};
      \draw (0,3) node[name=3] {{$\vdots$}};
      \draw (0,4) node[name=om] {{$\omega$}};
      \draw (0,5) node[name=om1] {{$\omega+1$}};
       }
      \draw[semithick] (0) -- (1);
      \draw[semithick] (1) -- (2);
      \draw[semithick] (om) -- (om1);
      
\end{tikzpicture}
}
\end{center}
\noindent
Let $f:C\ra C$ be defined as follows: 
  $$
  f(x) \ud \left\{
  	\begin{array}{ll}
	\omega & \mbox{ if $x<\omega$;} \\
	\omega+1 &\mbox{ otherwise.}
	\end{array}
	\right.
	$$
Let $\mu\in \uco(C)$ be the identity  uco $\lambda x.x$, 
so that $\mu \comp f \comp \mu = f$. For any $k\geq 0$, consider $\rho_k\in \uco(C)$ defined as 
$\ok{\rho_k \ud C\smallsetminus [0,k)}$. It is easily seen that, 
for any $k$, $\rho_k \comp f \comp \rho_k = f = \mu \comp f \comp \mu$. 
However, it turns out that $\sqcup_{k\geq 0} \rho_k = \cap_{k\geq 0} 
\img(\rho_k) =
\{\omega, \omega +1\}$, so that, for any $x\leq \omega$, 
$(\sqcup_{k\geq 0} \rho_k)(x)=\omega$, and $(\sqcup_{k\geq 0} \rho_k)(\omega +1)=\omega +1$.
It is then easy to check that 
$(\sqcup_{k\geq 0} \rho_k) \comp f \comp (\sqcup_{k\geq 0} \rho_k) = \lambda x.\omega +1\neq \mu \comp f \comp \mu$. As a consequence, the correctness kernel of $\mu$ for $f$ does not exist. Observe that $\mu \comp f=f$ is clearly not continuous and therefore this example is consistent with Theorem~\ref{kernel}.  
\qed
\end{example}

\section{Correctness Kernels in Abstract Model Checking}

\paragraph{\textbf{Partitioning Abstractions.}}
Following \cite{rt04,rt07}, 
partitions of a finite state space $\Sigma$ 
can be viewed as  abstractions
of the concrete domain $\wp(\Sigma)$.  
Let $\Part(\Sigma)$ denote the set of partitions of
$\Sigma$ and recall that $\tuple{\Part(\Sigma),\preceq,\curlyvee,\curlywedge}$ is 
a complete lattice, where $P_1 \preceq P_2$ iff for all $s\in \Sigma$, $P_1(s)\subseteq P_2(s)$.
Given a partition $P\in \Part(\Sigma)$, we consider 
the corresponding set $\wp(P)$ of all (possibly empty) sets of blocks of $P$.
Then, $\tuple{\wp(P),\subseteq}$ can be viewed as
an abstract domain of $\langle \wp(\Sigma),\subseteq\rangle$, 
which is called partitioning abstraction,
by means of the following Galois insertion
$(\alpha_P, \wp(\Sigma) , \wp(P), \gamma_P)$:
$$\alpha_P(X) \ud  \{B\in P~|~ B\cap X \neq \varnothing\}  
\text{~~and~~}
\gamma_P(\mathcal{B}) \ud \cup_{B\in \mathcal{B}}B.$$
Hence, the abstraction 
$\alpha_P(X)$ provides  the minimal over-approximation
of a set $X$ of states through blocks of $P$. 

Also, an abstraction $A\in \Abs(\wp(\Sigma))$ 
is called partitioning when there exists a partition $P\in \Part(\Sigma)$ such that 
$(\alpha_A,\wp(\Sigma),A,\gamma_A)$ is equivalent to $(\alpha_P, \wp(\Sigma) , \wp(P), \gamma_P)$. 
This happens exactly when $\gamma_A(A)\subseteq \wp(\Sigma)$ is closed under set intersections 
and complementations.

Finally, let us recall that any abstraction $A\in \Abs(\wp(\Sigma))$ induces a partition 
$P_A\in \Part(\Sigma)$ as follows: for any $s,t\in \Sigma$, 
$P_A(s)=P_A(t) \:\Lra\: \alpha_A(\{s\})=\alpha_A(\{t\})$. This is particularly interesting because
the corresponding partitioning abstraction $(\alpha_{P_A}, \wp(\Sigma) , \wp(P_A), \gamma_{P_A})$
turns out to be the least partitioning abstraction refinement of $A$. 

\paragraph{\textbf{Abstract Transition Systems.}}
Consider a finite state transition system $\mathcal{S} 
= \langle \Sigma, \sra \rangle$ and a corresponding 
abstract transition system $\cA =
\langle P, \sra^\sharp \rangle$ defined over a state
partition $P\in \Part(\Sigma)$. Equivalently, the abstract 
transition system $\cA$ could be defined over a set $A$ of abstract states
which is defined by a surjective function $h:\Sigma \ra A$ that induces a partition of $\Sigma$
(see e.g.\ \cite{cgp99}). 
Fixpoint-based verification of a  temporal specification
on the abstract model $\cA$ relies on 
computing some least/greatest
fixpoints of operators which are 
defined using Boolean connectives (union,
intersection, complementation) on abstract states and abstract
successor/predecessor functions $\post^\sharp$/$\pre^\sharp$
on the abstract transition system
$\langle P,\sra^\sharp\rangle$. 
The key point here is that
successor/predecessor functions are defined as best correct
approximations on the partitioning abstract domain $\wp(P)$
of the corresponding concrete successor/prede\-cessor functions on $\wp(\Sigma)$. 
In standard abstract model checking
\cite{bk08,cgl94,cgp99}, the abstract transition relation is defined
as the existential/existential relation $\sra^{\exists\exists}$
between blocks of $P$: for any $B,C\in P$,
$$ 
B \sra^{\exists\exists} C \text{~~~~iff~~~~} \exists x\in B.\exists y\in C.\: x
\sra y
$$
Accordingly, abstract predecessor and successor in 
$\tuple{P,\sra^{\exists\exists}}$ are given by the functions
$\pre^{\exists\exists}_P:\wp(P)\ra \wp(P)$
and $\post^{\exists\exists}_P:\wp(P)\ra \wp(P)$ defined as follows:
\begin{align*}
&\pre^{\exists\exists}_P(\mathcal{C}) \ud \{B\in P~|~\exists C\in \mathcal{C}.\,
B \sra^{\exists\exists} C\};\\
&\post^{\exists\exists}_P(\mathcal{B}) \ud \{C\in P~|~\exists B\in \mathcal{B}.\,
B \sra^{\exists\exists} C \}.
\end{align*}
As shown in \cite{rt04,rt07}, it turns out 
that $\pre^{\exists\exists}_P$
and $\post^{\exists\exists}_P$ are the best correct approximations of, respectively,
$\pre:\wp(\Sigma)\ra \wp(\Sigma)$ and $\post:\wp(\Sigma)\ra \wp(\Sigma)$  
on the abstraction $(\alpha_P, \wp(\Sigma) , \wp(P), \gamma_P)$.
In fact, for a set of blocks $\mathcal{C}\in \wp(P)$, we have that
\begin{align*}
\alpha_P(\pre(\gamma_P(\mathcal{C}))) &= \{B\in P~|~ B\cap \pre(\cup_{C\in \mathcal{C}} C)\neq \varnothing\}\\ 
& =\{B\in P~|~ \cup_{C\in \mathcal{C}} B\cap \pre(C)\neq \varnothing\}\\
& =\{B\in P~|~ \exists C\in \mathcal{C}.\: B \sra^{\exists\exists} C\}\\
&= \pre^{\exists\exists}_P(\mathcal{C})
\end{align*}
and analogous equations hold for $\post$. 
We thus have that
$$\pre^{\exists\exists}_P = \alpha_P \circ \pre \circ \gamma_P 
\text{~~and~~} \post^{\exists\exists}_P = \alpha_P \circ \post \circ \gamma_P.$$

\paragraph{\textbf{Correctness Kernels.}}
The above abstract interpretation-based approach
allows us to apply correctness kernels in abstract model checking as follows. 
The abstract transition system $\cA = \langle P, \sra^{\exists\exists} \rangle$ is 
viewed as an abstract interpretation which is defined 
by the abstract domain $(\alpha_P, \wp(\Sigma),$ 
$\wp(P), \gamma_P)$ and the abstract functions $\pre^{\exists\exists}_P = \alpha_P \circ \pre \circ \gamma_P$ and $\post^{\exists\exists}_P = \alpha_P \circ \post \circ \gamma_P$. We are thus interested in 
the correctness kernel of the partitioning abstraction $\wp(P)$ for
the concrete predecessor/successor functions $\{\pre,\post\}$, that we denote
simply by $\mathscr{K}_{\sra}(P)$. Observe that, 
by Theorem~\ref{kernel}, the kernel $\mathscr{K}_{\sra}(P)\in\Abs(\wp(\Sigma))$ 
clearly exists  since $\pre$, $\post$ and $\gamma_P\circ \alpha_P$  
are all additive functions on $\wp(\Sigma)$. The abstraction 
$\mathscr{K}_{\sra}(P)$ provides 
a simplification of the abstract domain 
$\wp(P)$ that preserves 
the best correct approximations of both predecessor and successor functions. 
In general, it turns out that 
$\mathscr{K}_{\sra}(P)$ is not a partitioning abstraction, as shown by the
following example.

\begin{example}\rm 
Consider the following 3-state transition system.
\begin{center}
    \begin{tikzpicture}[scale=0.40]
      \tikzstyle{arrow}=[->,>=latex']
      {\small 
      \path       
      (0,0) node[name=1]{$1$}
      (2,0) node[name=2]{$2$}
      (4,0) node[name=3]{$3$};
      }
      \draw[arrow,shorten >=-3pt, shorten <=-3pt] (1) to (2);
      \draw[arrow,shorten >=-3pt, shorten <=-3pt] (2) to (3);
      \draw[arrow,shorten >=-3pt, shorten <=-3pt] (1) to[bend left] (3);
\end{tikzpicture}
\end{center}
Let us consider the finest partition $P=\{[1],[2],[3]\}$, so that 
$\pre^{\exists\exists}_P = \pre$ and $\post^{\exists\exists}_P = \post$. 
In order to apply Theorem~\ref{kernel}, here we have that
$\img(\pre) = \{\varnothing, [1], [1,2]\}$,  
$\img(\post) = \{\varnothing, [3], [2,3]\}$ and
\begin{align*}
\max(\{S\in \wp(\Sigma)~|~\pre(S)=\varnothing\})&=\{1\}\\ 
\max(\{S\in \wp(\Sigma)~|~\pre(S)=\{1\}\})&=\{1,2\}\\ 
\max(\{S\in \wp(\Sigma)~|~\pre(S)=\{1,2\}\})&=\{1,2,3\}\\ 
\max(\{S\in \wp(\Sigma)~|~\post(S)=\varnothing\})&=\{3\}\\ 
\max(\{S\in \wp(\Sigma)~|~\post(S)=\{3\}\})&=\{2,3\}\\ 
\max(\{S\in \wp(\Sigma)~|~\post(S)=\{2,3\}\})&=\{1,2,3\}\\ 
\end{align*} 
Thus, by Theorem~\ref{kernel}, 
$$\mathscr{K}_{\sra}(P) = \Cl_{\cap}(\{\varnothing, [1],[3],[1,2],[2,3],[1,2,3]\})=
\{\varnothing, [1],[2],[3],[1,2],[2,3],[1,2,3]\}.$$ 
It turns out that $\mathscr{K}_{\sra}(P)$ is not partitioning, since it is not closed
under set unions. \qed
\end{example}

Since the abstract domain $\mathscr{K}_{\sra}(P)$ in general is not partitioning, we are thus
interested in its partitioning abstraction, which is characterized as follows.

\begin{corollary}\label{kernel-part}
Let $B_1,B_2 \in P$. Then,
$\mathscr{K}_{\sra}(P) (B_1) = \mathscr{K}_{\sra}(P) (B_2)$ 
if and only if  $\pre^{\exists\exists}_P (\{B_1\}) = 
\pre^{\exists\exists}_P (\{B_2\})$ and 
$\post^{\exists\exists}_P (\{B_1\}) = 
\post^{\exists\exists}_P (\{B_2\})$.
\end{corollary}
\begin{proof}
The kernel  $\mathscr{K}_{\sra}(P)$ can be obtained by 
applying Theorem~\ref{kernel} to the abstraction $(\alpha_P,\wp(\Sigma),\wp(P),\gamma_P)$ 
and to the functions $\{\pre,\post\}$. Since the best correct approximations 
$\pre^{\exists\exists}_P$ and $\post^{\exists\exists}_P$ are additive functions
on $\tuple{\wp(P),\subseteq}$, $\max$'s can be replaced by lub's in
$\tuple{\wp(P),\subseteq}$, namely set unions. Hence, we have that:
\begin{multline*}
\mathscr{K}_{\sra}(P) = \Cl_{\cap} 
\Big(\img(\pre^{\exists\exists_P}) \:{\textstyle \bigcup} \:
\big\{\!\cup\!\{\mathcal{C}\in \wp(P)~|~ \pre^{\exists\exists}(\mathcal{C}) = \mathcal{B}\}
~\big|~ \mathcal{B}\in \img(\pre^{\exists\exists}_P)
\big\}\\
{\bigcup} 
\img(\post^{\exists\exists}_P) \:{\textstyle \bigcup}\:
\big\{\!\cup \!\{\mathcal{B}\in \wp(P)~|~ \post^{\exists\exists}(\mathcal{B}) = \mathcal{C}\}
~\big|~ \mathcal{C}\in \img(\post^{\exists\exists}_P)
\big\}
\Big).
\end{multline*}

\noindent
Let us then show the stated equivalence. 

\medskip
\noindent
$(\Ra)$ More in general, it is enough to observe that 
if $\rho$ is the correctness kernel of some $\mu$ for some $f$ then for
any $c_1,c_2\in C$ such that $\rho(c_1)=\rho(c_2)$ we have that 
$\mu f \mu (c_1) = \rho f \rho (c_1) = \rho f \rho(c_2) = \mu f \mu (c_2)$.

\medskip
\noindent
$(\La)$ In the following, let $\mu \in \uco(\wp(\Sigma))$ denote the uco induced by the
abstraction $\mathscr{K}_{\sra}(P)$. 
Let us consider two blocks $B_1,B_2\in P$.
If 
$\pre^{\exists\exists}_P (\mathcal{C}) \in \img(\pre^{\exists\exists}_P)$, for some
$\mathcal{C}\in \wp(P)$, then we have that:
\begin{align*}
B_1 \in \pre^{\exists\exists}_P (\mathcal{C}) 
& \Leftrightarrow 
\text{~~~~\big[by definition of $\pre^{\exists\exists}_P$\big]}
\\
\exists C\in \mathcal{C}.\,  
B_1 \sra^{\exists\exists} C  & \Leftrightarrow 
\text{~~~~\big[by definition of $\post^{\exists\exists}_P$\big]}\\
\exists C\in \mathcal{C}.\,  
C \in \post^{\exists\exists}_P (\{B_1\}) & \Leftrightarrow 
\text{~~~~\big[by hypothesis\big]}\\
\exists C\in \mathcal{C}.\,  
C \in \post^{\exists\exists}_P (\{B_2\}) & \Leftrightarrow 
\text{~~~~\big[by replicating the previous arguments\big]}\\
B_2 \in \pre^{\exists\exists}_P (\mathcal{C})  &
\end{align*}

\noindent
Also, for any $\mathcal{B}\in \img(\pre^{\exists\exists}_P)$, we have that:
 \begin{align*}
B_1 \in \cup\{\mathcal{C}\in \wp(P)~|~ 
\pre^{\exists\exists}_P(\mathcal{C}) = \mathcal{B}\}
& \Leftrightarrow 
\text{~~~~\big[by definition of $\pre^{\exists\exists}_P$\big]}\\
B_1 \in \{C\in P~|~ \pre^{\exists\exists}_P(\{C\}) \subseteq \mathcal{B}\} 
& \Leftrightarrow 
\\
\pre^{\exists\exists}_P(\{B_1\}) \subseteq \mathcal{B} 
& \Leftrightarrow 
\text{~~~~\big[by hypothesis\big]}\\
\pre^{\exists\exists}_P(\{B_2\}) \subseteq \mathcal{B} 
& \Leftrightarrow 
\text{~~~~\big[by replicating the previous arguments\big]}\\
B_2 \in \cup\{\mathcal{C}\in \wp(P)~|~ 
\pre^{\exists\exists}_P(\mathcal{C}) = \mathcal{B}\}  &
\end{align*}

\noindent
Likewise, for any $\mathcal{B}\in \wp(P)$ and 
$\mathcal{C}\in \img(\post^{\exists\exists}_P)$ we also have that:
\begin{align*}
&B_1 \in \post^{\exists\exists}_P (\mathcal{B})\Lra B_2 \in \post^{\exists\exists}_P (\mathcal{B})\\
&B_1 \in \cup\{\mathcal{B}\in \wp(P)~|~ 
\post^{\exists\exists}_P(\mathcal{B}) = \mathcal{C}\} \Lra
B_2 \in \cup\{\mathcal{B}\in \wp(P)~|~ 
\post^{\exists\exists}_P(\mathcal{B}) = \mathcal{C}\}.
\end{align*}

\noindent
Consequently, $\mathscr{K}_{\sra}(P)(B_1) = \mathscr{K}_{\sra}(P)(B_2)$. 
\end{proof}

We denote by $P_\cK\in \Part(\Sigma)$ the partitioning abstraction of $\mathscr{K}_{\sra}(P)$. 
We therefore have that in $P_\cK$ a block $B\in P$ is merged together with all 
the blocks $B'\in P$ such that for any block $B'\in P$, 
$\pre^{\exists\exists} (\{B\}) = \pre^{\exists\exists} (\{B'\})$ and 
$\post^{\exists\exists} (\{B\}) = \post^{\exists\exists} (\{B'\})$. 

Given $P,Q\in \Part(\Sigma)$, let $\pre^{\exists\exists}_Q = \pre^{\exists\exists}_P$
denote the fact that for all $s\in \Sigma$, $\cup \pre^{\exists\exists}_Q (Q(s))=
\cup \pre^{\exists\exists}_P (P(s))$, and analogously for $\post$. We thus derive the following
characterization of $P_\cK$.

\begin{corollary}\label{coronew}
$P_\cK = \curlyvee \{Q\in \Part(\Sigma)~|~\pre^{\exists\exists}_Q =
\pre^{\exists\exists}_P ,\:\post^{\exists\exists}_Q =
\post^{\exists\exists}_P \}$.
\end{corollary}
\begin{proof}
Let $\mu \ud \mathscr{K}_{\sra}(P)\in \Abs(\wp(\Sigma))$. 
Let us first check that $\pre^{\exists\exists}_{P_\cK} =
\pre^{\exists\exists}_P$. Given $s\in \Sigma$, we have that:
\begin{align*}
\cup \pre^{\exists\exists}_P (P(s)) & \subseteq 
\text{~~~~\big[since $P\preceq P_\cK$\big]}\\
\cup \pre^{\exists\exists}_{P_\cK} (P_\cK(s)) & \subseteq
\text{~~~~\big[since, for any $S$, $P_\cK(S) \subseteq \mu(S)$\big]}\\
\mu(\pre (\mu(s))) & = 
\text{~~~~\big[since $\mu$ is the correctness kernel of $P$ for $\pre$ and $\post$\big]}\\
\cup \pre^{\exists\exists}_P (P(s)) & = 
\end{align*}
Hence, $\pre^{\exists\exists}_{P_\cK} =
\pre^{\exists\exists}_P$. Likewise, $\post^{\exists\exists}_{P_\cK} =
\post^{\exists\exists}_P$. Therefore, we obtain that $P_\cK \preceq \curlyvee \{Q\in \Part(\Sigma)~|~\pre^{\exists\exists}_Q =
\pre^{\exists\exists}_P ,\post^{\exists\exists}_Q =
\post^{\exists\exists}_P \}$. On the other hand, if $Q\in \Part(\Sigma)$ is such that 
$\pre^{\exists\exists}_Q = \pre^{\exists\exists}_P$ and 
$\post^{\exists\exists}_Q = \post^{\exists\exists}_P$ then, since 
$\mu$ is the correctness kernel of $P$ for $\pre$ and $\post$, $\gamma_Q \circ \alpha_Q \sqsubseteq 
\mu$. Hence, since the partitioning abstraction refinement is monotonic, we obtain that
$Q\preceq P_\cK$. Consequently, $\{Q\in \Part(\Sigma)~|~\pre^{\exists\exists}_Q =
\pre^{\exists\exists}_P ,\post^{\exists\exists}_Q =
\post^{\exists\exists}_P \}\preceq P_\cK$.
\end{proof}

\begin{example}\rm
Reconsider the abstract transition system $\cA$ in Figure~\ref{figure-1}
where the underlying state partition is 
$P=\{[1], [2,3],$ $[4,5], [6], [7], [8,9]\}$.
 Here, by Corollary~\ref{kernel-part},
 the block 
 $[2,3]$ is merged with $[4,5]$ while $[6]$ is merged with $[7]$. This therefore 
 simplifies 
 the partition $P$ to $P_\cK= \{[1],$ $[2,3,4,5], [6,7], [8,9]\}$, that is, we obtain
 the abstract transition system $\cA''$ depicted in Figure~\ref{figure-1}. 
\qed
\end{example}

\section{Example Guided Abstraction Simplification}\label{egas}
Let us discuss how correctness kernels 
give rise to an Example-Guided Abstraction Simplification (EGAS) paradigm
in abstract transition systems. 

\subsection{CEGAR Background}
Let us first recall some basic notions of 
CEGAR \cite{cgjlv00,cgjlv03}. 
Consider an abstract transition system $\cA =
\langle P, \sra^{\exists\exists}\rangle$ defined over a state
partition $P\in \Part(\Sigma)$ and some finite abstract path $\pi =
\langle B_1,...,B_n\rangle$
in $\cA$, where each $B_i$ is a block of $P$. 
Typically, this path is a counterexample to the validity in $\cA$ of a temporal
formula and it originated as  output of a model checker 
running on $\cA$ (for simplicity we do not consider here loop path counterexamples). 
The set of concrete paths that are abstracted to $\pi$ are defined as follows:
$$\paths(\pi) \ud \{\tuple{s_1,...,s_n}\in \Sigma^n~|~\forall i\in [1,n]. 
s_i\in B_i \;\&\; \forall i\in [1,n). s_i \sra s_{i+1}\}.$$ 
The abstract path $\pi$
is \emph{spurious} when it represents no real concrete path, that is, 
when $\paths(\pi)=\varnothing$. A corresponding 
sequence $\spu(\pi) = 
\langle S_1,...,S_n\rangle$
of sets of states in $\Sigma$ is  inductively 
defined as follows: $S_1 \ud B_1$; $S_{i+1} \ud \post(S_i)\cap B_{i+1}$. 
As observed in \cite{cgjlv03}, it turns out that $\pi$ is spurious iff there exists 
a least $k\in [1,n-1]$ such that $S_{k+1}=\varnothing$. In such a case,
the partition $P$ is refined by splitting the block $B_{k}$. The 
three following sets partition the states of the block $B_{k}$:

\medskip
dead-end states: $B_{k}^{\text{dead}} \ud S_{k}\neq \varnothing$

\smallskip
bad states: $B_{k}^{\text{bad}} \ud B_{k} \cap \pre(B_{k+1}) \neq \varnothing$

\smallskip
irrelevant states: $B_{k}^{\text{irr}} \ud B_{k}\smallsetminus
(B_{k}^{\text{dead}} \cup B_{k}^{\text{bad}})$

\medskip
\noindent
The split of the block $B_{k}$ 
must separate dead-end states from bad states, while irrelevant states may be joined indifferently with dead-end or bad states. 
However, when states are memory stores, 
the problem of finding the coarsest refinement of $P$ that separates
dead-end and bad states is NP-hard~\cite[Theorem~4.17]{cgjlv03} and thus 
some refinement heuristics are necessarily used. According to the
basic heuristics of CEGAR \cite[Section~4]{cgjlv03}, $B_{k}$ is simply split into 
$B_{k}^{\text{dead}}$ and $B_{k}^{\text{bad}} \cup B_{k}^{\text{irr}}$. 

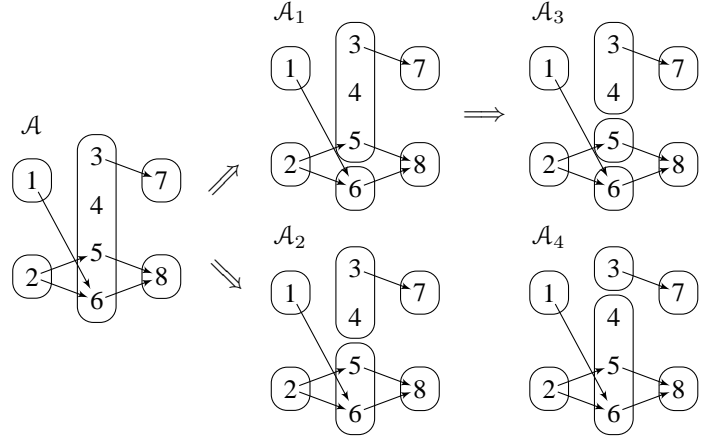
\begin{figure}[t]
\begin{center}
\normalsize
    \begin{tikzpicture}[scale=0.425,shorten >=-3pt, shorten <=-3pt]
      \tikzstyle{arrow}=[->,>=latex']
      \path 
      (0,2.25) node[name=1]{1}
      (0,-0.75) node[name=2]{2}
      (2,3) node[name=3]{3}
      (2,1.5) node[name=4]{4}
      (2,0) node[name=5]{5}
      (2,-1.5) node[name=6]{6}
      (4,2.25) node[name=7]{7}
      (4,-0.75) node[name=8]{8};
      \path (6,2.25) node[name=f,rotate=45]{$\Longrightarrow$};
      \path (6,-0.75) node[name=f,rotate=-45]{$\Longrightarrow$};

      \path (0,4) node[name=a]{$\mathcal{A}$};
      \draw[arrow] (1) to (6);
      \draw[arrow] (2) to (5);
      \draw[arrow] (2) to (6);
      \draw[arrow] (3) to (7);
      \draw[arrow] (5) to (8);
      \draw[arrow] (6) to (8);
                      
    \path (1.north west) ++(-0.1,0.1) node[name=a1]{} (1.south east) ++(0.1,-0.1) 
      node[name=a2]{};
     \draw[rounded corners=6pt] (a1) rectangle (a2);

    \path (2.north west) ++(-0.1,0.1) node[name=a3]{} (2.south east) ++(0.1,-0.1) 
      node[name=a4]{};
     \draw[rounded corners=6pt] (a3) rectangle (a4);

     \path (3.north west) ++(-0.1,0.1) node[name=b3]{} (6.south east) ++(0.1,-0.1) 
      node[name=b4]{};
     \draw[rounded corners=6pt] (b3) rectangle (b4);

     \path (7.north west) ++(-0.1,0.1) node[name=c1]{} (7.south east) ++(0.1,-0.1) 
      node[name=c2]{};
     \draw[rounded corners=6pt] (c1) rectangle (c2);

	\path (8.north west) ++(-0.1,0.1) node[name=c3]{} (8.south east) ++(0.1,-0.1) 
      node[name=c4]{};
     \draw[rounded corners=6pt] (c3) rectangle (c4);

\path 
      (8,5.75) node[name=1]{1}
      (8,2.75) node[name=2]{2}
      (10,6.5) node[name=3]{3}
      (10,5) node[name=4]{4}
      (10,3.5) node[name=5]{5}
      (10,2) node[name=6]{6}
      (12,5.75) node[name=7]{7}
      (12,2.75) node[name=8]{8};
      \path (14,4.25) node[name=f]{$\Longrightarrow$};
      \path (8,7.5) node[name=a]{$\mathcal{A}_1$};

      \draw[arrow] (1) to (6);
      \draw[arrow] (2) to (5);
      \draw[arrow] (2) to (6);
      \draw[arrow] (3) to (7);
      \draw[arrow] (5) to (8);
      \draw[arrow] (6) to (8);
  
    \path (1.north west) ++(-0.1,0.1) node[name=a1]{} (1.south east) ++(0.1,-0.1) 
      node[name=a2]{};
     \draw[rounded corners=6pt] (a1) rectangle (a2);

    \path (2.north west) ++(-0.1,0.1) node[name=a3]{} (2.south east) ++(0.1,-0.1) 
      node[name=a4]{};
     \draw[rounded corners=6pt] (a3) rectangle (a4);

     \path (3.north west) ++(-0.1,0.1) node[name=b3]{} (5.south east) ++(0.1,-0.1) 
      node[name=b4]{};
     \draw[rounded corners=6pt] (b3) rectangle (b4);

     \path (6.north west) ++(-0.1,0.1) node[name=b5]{} (6.south east) ++(0.1,-0.1) 
      node[name=b6]{};
     \draw[rounded corners=6pt] (b5) rectangle (b6);

     \path (7.north west) ++(-0.1,0.1) node[name=c1]{} (7.south east) ++(0.1,-0.1) 
      node[name=c2]{};
     \draw[rounded corners=6pt] (c1) rectangle (c2);

	\path (8.north west) ++(-0.1,0.1) node[name=c3]{} (8.south east) ++(0.1,-0.1) 
      node[name=c4]{};
     \draw[rounded corners=6pt] (c3) rectangle (c4);

   \path 
      (8,-1.25) node[name=1]{1}
      (8,-4.25) node[name=2]{2}
      (10,-0.5) node[name=3]{3}
      (10,-2) node[name=4]{4}
      (10,-3.5) node[name=5]{5}
      (10,-5) node[name=6]{6}
      (12,-1.25) node[name=7]{7}
      (12,-4.25) node[name=8]{8};
      \path (8,0.5) node[name=a]{$\mathcal{A}_2$};

      \draw[arrow] (1) to (6);
      \draw[arrow] (2) to (5);
      \draw[arrow] (2) to (6);
      \draw[arrow] (3) to (7);
      \draw[arrow] (5) to (8);
      \draw[arrow] (6) to (8);
  
    \path (1.north west) ++(-0.1,0.1) node[name=a1]{} (1.south east) ++(0.1,-0.1) 
      node[name=a2]{};
     \draw[rounded corners=6pt] (a1) rectangle (a2);

    \path (2.north west) ++(-0.1,0.1) node[name=a3]{} (2.south east) ++(0.1,-0.1) 
      node[name=a4]{};
     \draw[rounded corners=6pt] (a3) rectangle (a4);

     \path (3.north west) ++(-0.1,0.1) node[name=b3]{} (4.south east) ++(0.1,-0.1) 
      node[name=b4]{};
     \draw[rounded corners=6pt] (b3) rectangle (b4);

     \path (5.north west) ++(-0.1,0.1) node[name=b5]{} (6.south east) ++(0.1,-0.1) 
      node[name=b6]{};
     \draw[rounded corners=6pt] (b5) rectangle (b6);

     \path (7.north west) ++(-0.1,0.1) node[name=c1]{} (7.south east) ++(0.1,-0.1) 
      node[name=c2]{};
     \draw[rounded corners=6pt] (c1) rectangle (c2);

	\path (8.north west) ++(-0.1,0.1) node[name=c3]{} (8.south east) ++(0.1,-0.1) 
      node[name=c4]{};
     \draw[rounded corners=6pt] (c3) rectangle (c4);

   \path 
      (16,5.75) node[name=1]{1}
      (16,2.75) node[name=2]{2}
      (18,6.5) node[name=3]{3}
      (18,5) node[name=4]{4}
      (18,3.5) node[name=5]{5}
      (18,2) node[name=6]{6}
      (20,5.75) node[name=7]{7}
      (20,2.75) node[name=8]{8};
      
      \path (16,7.5) node[name=a]{$\mathcal{A}_3$};
      \draw[arrow] (1) to (6);
      \draw[arrow] (2) to (5);
      \draw[arrow] (2) to (6);
      \draw[arrow] (3) to (7);
      \draw[arrow] (5) to (8);
      \draw[arrow] (6) to (8);
                      
    \path (1.north west) ++(-0.1,0.1) node[name=a1]{} (1.south east) ++(0.1,-0.1) 
      node[name=a2]{};
     \draw[rounded corners=6pt] (a1) rectangle (a2);

    \path (2.north west) ++(-0.1,0.1) node[name=a3]{} (2.south east) ++(0.1,-0.1) 
      node[name=a4]{};
     \draw[rounded corners=6pt] (a3) rectangle (a4);

     \path (3.north west) ++(-0.1,0.1) node[name=b3]{} (4.south east) ++(0.1,-0.1) 
      node[name=b4]{};
     \draw[rounded corners=6pt] (b3) rectangle (b4);

    \path (5.north west) ++(-0.1,0.1) node[name=bb3]{} (5.south east) ++(0.1,-0.1) 
      node[name=bb4]{};
     \draw[rounded corners=6pt] (bb3) rectangle (bb4);

  \path (6.north west) ++(-0.1,0.1) node[name=db3]{} (6.south east) ++(0.1,-0.1) 
      node[name=db4]{};
     \draw[rounded corners=6pt] (db3) rectangle (db4);

     \path (7.north west) ++(-0.1,0.1) node[name=c1]{} (7.south east) ++(0.1,-0.1) 
      node[name=c2]{};
     \draw[rounded corners=6pt] (c1) rectangle (c2);

	\path (8.north west) ++(-0.1,0.1) node[name=c3]{} (8.south east) ++(0.1,-0.1) 
      node[name=c4]{};
     \draw[rounded corners=6pt] (c3) rectangle (c4);

\path 
      (16,-1.25) node[name=1]{1}
      (16,-4.25) node[name=2]{2}
      (18,-0.5) node[name=3]{3}
      (18,-2) node[name=4]{4}
      (18,-3.5) node[name=5]{5}
      (18,-5) node[name=6]{6}
      (20,-1.25) node[name=7]{7}
      (20,-4.25) node[name=8]{8};
      
      \path (16,0.5) node[name=a]{$\mathcal{A}_4$};
      \draw[arrow] (1) to (6);
      \draw[arrow] (2) to (5);
      \draw[arrow] (2) to (6);
      \draw[arrow] (3) to (7);
      \draw[arrow] (5) to (8);
      \draw[arrow] (6) to (8);
                      
    \path (1.north west) ++(-0.1,0.1) node[name=a1]{} (1.south east) ++(0.1,-0.1) 
      node[name=a2]{};
     \draw[rounded corners=6pt] (a1) rectangle (a2);

    \path (2.north west) ++(-0.1,0.1) node[name=a3]{} (2.south east) ++(0.1,-0.1) 
      node[name=a4]{};
     \draw[rounded corners=6pt] (a3) rectangle (a4);

     \path (3.north west) ++(-0.1,0.1) node[name=b3]{} (3.south east) ++(0.1,-0.1) 
      node[name=b4]{};
     \draw[rounded corners=6pt] (b3) rectangle (b4);

    \path (4.north west) ++(-0.1,0.1) node[name=bb3]{} (6.south east) ++(0.1,-0.1) 
      node[name=bb4]{};
     \draw[rounded corners=6pt] (bb3) rectangle (bb4);

     \path (7.north west) ++(-0.1,0.1) node[name=c1]{} (7.south east) ++(0.1,-0.1) 
      node[name=c2]{};
     \draw[rounded corners=6pt] (c1) rectangle (c2);

	\path (8.north west) ++(-0.1,0.1) node[name=c3]{} (8.south east) ++(0.1,-0.1) 
      node[name=c4]{};
     \draw[rounded corners=6pt] (c3) rectangle (c4);

  \end{tikzpicture}    
\end{center}
\caption{Some abstract transition systems.}\label{figura-bis}
\end{figure}
Let us see a simple example. 
Consider the abstract path $\pi=\langle[1], [3,4,5,6], [7]\rangle$ in the
abstract transition system $\cA$ depicted in Figure~\ref{figura-bis}. 
This is a spurious path and the block $[3,4,5,6]$ needs to be split. 
This block is therefore partitioned as follows:
$[6]$ dead-end states, $[3]$ bad states and $[4,5]$ irrelevant states.
The refinement heuristics of CEGAR tells us that irrelevant states are joined
with bad states 
so that $\cA$ is refined to the abstract transition system $\cA_1$. In turn, consider the
spurious path $\pi' = \langle [2], [3,4,5], [7] \rangle$ in $\cA_1$, 
so that CEGAR refines $\cA_1$ to $\cA_3$ by splitting the block $[3,4,5]$ into 
$[3,4]$ and $[5]$, i.e.,  
bad and irrelevant states in $[3,4]$ and dead-end states in $[5]$.    
In the first abstraction refinement, let us observe that if 
irrelevant states in $[4,5]$ 
would have been joined together with dead-end states in $[6]$ 
rather than with bad
states in $[3]$ we would have obtained the abstract system $\cA_4$, and
$\cA_4$ does not contain spurious paths so that it does
not need to be further refined. Let us also notice that 
if the irrelevant state $5$ would have been joined
with dead-end states $[6]$ while the irrelevant state $4$ would have been joined
with bad states $[3]$ we would have obtained the abstract system $\cA_2$ that 
still does not need to be further refined since it does not contain spurious paths. 

\subsection{EGAS}
EGAS can be integrated within
the CEGAR loop thanks to the following remark. 
If $\pi_1$ and $\pi_2$ are paths, respectively, 
in $\tuple{P_1,\sra^{\exists\exists}}$ and $\tuple{P_2,\sra^{\exists\exists}}$,
where 
$P_1,P_2\in \Part(\Sigma)$ and $P_1\preceq P_2$, then we say that 
$\pi_1$ is abstracted to $\pi_2$, denoted by
$\pi_1 \sqsubseteq \pi_2$,
when $\length(\pi_1)=\length(\pi_2)$
and for any $j\in [1,\length(\pi_1)]$, $\pi_1(j)\subseteq \pi_2(j)$. 

\begin{corollary}\label{coro2}
Consider an abstract transition system $\cA=\tuple{P,\sra^{\exists\exists}}$ 
over a partition $P\in \Part(\Sigma)$ and its simplification 
$\cA_s=\tuple{P_\cK,\sra^{\exists\exists}}$ induced by the correctness
kernel $\mathscr{K}_{\sra}(P)$. If $\pi$ is a spurious abstract path
in $\cA_s$ then there exists a spurious abstract path $\pi'$ in $\cA$ 
such that $\pi'\sqsubseteq \pi$.  
\end{corollary}
\begin{proof}
Let 
$\pi = \langle B_1,...,B_n\rangle$, where, for any $i\in [1,n]$, $B_i \in P_\cK$, 
and let $B_k$ be the block of $\pi$ that 
generates the spuriousness of $\pi$. 
Since $P \preceq P_\cK$, 
we have that for each $i\in [1,n]$, $B_i = \cup_{j_i \in J_i} C_i^{j_i}$, for some set of 
blocks $C_i^{j_i} \in P$. 
By Corollary~\ref{coronew}, for each $i\in [1,n)$ and $j_i\in J_i$, 
$\cup \post^{\exists\exists}_{P_\cK} 
(B_i) = \cup \post^{\exists\exists}_P (C_i^{j_i})$ and 
for each $i\in (1,n]$ and $j_i\in J_i$, $\cup\pre^{\exists\exists}_{P_\cK} 
(B_i) = \cup\pre^{\exists\exists}_P (C_i^{j_i})$. Then, in order to define the path $\pi'$ in $\cA$, for 
any $i \in [1,n]$, 
one can choose any block $C_i^{j_i}$ in $P$ such that $C_i^{j_i} \subseteq B_i$. 
The key point to note is that by Corollary~\ref{coronew}, it turns out that
$C_k^{j_k}$ causes the spuriousness of the path $\pi'$. Moreover, $\pi' \sqsubseteq \pi$, and
this concludes the proof.
\end{proof}

This means  that the abstraction simplification induced by 
the correctness kernel does not add spurious paths. 

\subsection{Bad- and Dead-irrelevant States}
The above observations suggest 
us a new refinement strategy within the CEGAR loop.
Let $\pi = \langle B_1,...,B_n\rangle$
be a spurious path in $\cA$ and let $\spu(\pi) = 
\langle S_1,...,S_n\rangle$ such that $S_{k+1}=\varnothing$ for some minimum
$k\in [1,n-1]$, so that the block $B_{k}$ needs to
be split. The set of irrelevant states in $B_{k}^{\text{irr}}$ is thus 
partitioned as specified by the following strategy.
An irrelevant state $s\in B_{k}^{\text{irr}}$
is called \emph{bad-irrelevant} when 
$$
\pre^{\exists\exists}_P(B_{k}^{\text{bad}} \cup \{s\}) = 
\pre^{\exists\exists}_P(B_{k}^{\text{bad}}) \text{~~and~~} 
\post^{\exists\exists}_P(B_{k}^{\text{bad}} \cup \{s\}) = 
\post^{\exists\exists}_P(B_{k}^{\text{bad}})
$$

\noindent
Thus, a bad-irrelevant state 
can be joined to bad states without affecting the set of abstract paths in $P$
that go through  $B_{k}^{\text{bad}}$. 
\emph{Dead-irrelevant} states  are analogously defined w.r.t.\ 
the set $B_{k}^{\text{dead}}$ of dead-end states. 
It may happen that an irrelevant state $s$ is both bad- and dead-irrelevant: 
in this case, $s$ could be equivalently merged with bad or dead states since
in both cases no spurious path would be added. Clearly, it may also
happen that an irrelevant state  is neither bad- nor dead-irrelevant. 
These states are called \emph{fully-irrelevant}. 

Let us denote by
$S_{k}^{\text{bad-irr}}$ and $S_{k}^{\text{dead-irr}}$, respectively, 
the set of all bad- and dead-irrelevant states 
in $B_{k}^{\text{irr}}$. 
We can therefore partition the set of irrelevant states in $B_{k}^{\text{irr}}$ 
as follows: 

\bigskip
bad-irrelevant block: 
$B_{k}^{\text{bad-irr}} \ud  S_{k}^{\text{bad-irr}} \smallsetminus S_{k}^{\text{dead-irr}}$

\smallskip
dead-irrelevant block: $B_{k}^{\text{dead-irr}} \ud  S_{k}^{\text{dead-irr}} \smallsetminus S_{k}^{\text{bad-irr}}$

\smallskip
fully-irrelevant block: 
$B_{k}^{\text{fully-irr}} \ud (S_{k}^{\text{bad-irr}} \cap S_{k}^{\text{dead-irr}}) 
\cup \big(B_{k}^{\text{irr}}\smallsetminus (S_{k}^{\text{bad-irr}} \cup S_{k}^{\text{dead-irr}})\big)$

\bigskip
Hence, the set of irrelevant states in $B_{k}^{\text{irr}}$ 
is partitioned into three disjoint blocks: $B_{k}^{\text{bad-irr}}$, 
$B_{k}^{\text{dead-irr}}$ and $B_{k}^{\text{fully-irr}}$. Notice that it may
happen that one or two of these sets is empty, whereas at least one of them must
be non-empty. 

We denote by $P^{\pi}$ the refined partition obtained from $P$ by replacing the 
block $B_k$ with at most five (and at least three) non-empty blocks: 
$B_{k}^{\text{bad}}$, 
$B_{k}^{\text{bad-irr}}\neq \varnothing$, $B_{k}^{\text{dead}}$, 
$B_{k}^{\text{dead-irr}}\neq \varnothing$ 
and $B_{k}^{\text{fully-irr}}\neq \varnothing$. 
By Corollary~\ref{coronew}, it is clear that in the partition  
$P^{\pi}_\cK$ obtained from the correctness kernel 
of $P^{\pi}$,  $B_{k}^{\text{bad}}$ is merged with
$B_{k}^{\text{bad-irr}}$,  $B_{k}^{\text{dead}}$ is merged with
$B_{k}^{\text{dead-irr}}$, while $B_{k}^{\text{fully-irr}}$ remains 
a separate block in $\mathscr{K}_{\sra}(P_{\pi})$. 
Also, by Corollary~\ref{coro2}, it turns out that 
no spurious path is added in the abstract system 
$\langle P^{\pi}_\cK,\rightarrow^{\exists\exists}  \rangle$ 
w.r.t.\ the system $\langle P^{\pi},\rightarrow^{\exists\exists}  \rangle$.  

Summing up, the refinement strategy  EGAS goes as follows:   
\begin{itemize}
\item[(A)] If $B_{k}^{\text{bad-irr}}\neq \varnothing$ then 
merge $B_{k}^{\text{bad-irr}}$ with bad states.
\item[(B)] If $B_{k}^{\text{dead-irr}}\neq \varnothing$ then 
merge $B_{k}^{\text{dead-irr}}$ with dead-end states.
\item[(C)] If $B_{k}^{\text{fully-irr}}\neq\varnothing$ then these 
fully-irrelevant states 
can be indifferently merged with bad or dead states;
for these states, one could use, for example, 
the basic refinement heuristics of CEGAR that merge them with bad states. 
\end{itemize}

In the above example, for the spurious path 
$\tuple{[1],[3,4,5,6],[7]}$ in $\cA$, the block $B=[3,4,5,6]$
needs to be refined. We have that:
$$
B^{\text{bad}}=[3],\; B^{\text{dead}}=[6],\; 
B^{\text{irr}}=[4,5].
$$
 Here, $5$ is a dead-irrelevant
state because $\pre^{\exists\exists}([5,6]) = \{[1],[2]\} = \pre^{\exists\exists}([6])$ and
$\post^{\exists\exists}([5,6])= \{[8]\}=\post^{\exists\exists}([6])$; also,
$5$ is not bad-irrelevant because 
$\pre^{\exists\exists}([3,5]) \neq \pre^{\exists\exists}([3])$.
Moreover, $4$ is both dead- and bad-irrelevant and therefore it is fully-irrelevant. 
Hence, according to the EGAS refinement strategy, 
the block $[3,4,5,6]$ is split into 
$[3,4]$ and $[5,6]$, so that EGAS gives rise to the abstract system $\cA_2$ that
does not need further refinements.   

\section{Correctness Kernels in Predicate Abstraction}
Let us discuss how 
correctness kernels can be also used in the context of predicate abstraction-based  
model checking \cite{ddp99,gs97}. Following Ball et al.'s approach \cite{bpr03}, 
predicate abstraction can be formalized by abstract interpretation as follows. 
Let us consider a program $P$ with $k$ integer variables $x_1$,...,$x_k$. 
The concrete domain of computation of $P$ is
$\tuple{\wp(\States),\subseteq}$ where 
$\States \ud \{x_1,...,x_k\} \rightarrow \mathbb{Z}$. 
Values in $\States$ are denoted by tuples $\tuple{z_1,...,z_k}\in \mathbb{Z}^k$. 
The program $P$ generates a transition system 
$\tuple{\States,\sra}$ so that the 
concrete semantics of $P$ is defined by the corresponding 
successor function
$\post:\wp(\States) \ra \wp(\States)$. 

A finite set $\mathcal{P} = \{p_1,...,p_n\}$ of state predicates is considered, 
where each predicate $p_i$ denotes the subset of states that satisfy $p_i$,
i.e.\ $\{s\in \States ~|~ s \models p_i\}$.  
These predicates give rise to 
the so-called \emph{Boolean abstraction} 
$B \ud \langle \wp(\{0,1\}^n),\subseteq \rangle$ 
which is related to $\wp(\States)$ through the following abstraction and concretization
maps (here, $s\models p_i$ is understood to assume values in $\{0,1\}$):
\begin{align*}
\alpha_B(S) &\ud \{\langle s\models p_1,...,s\models p_n \rangle \in 
\{0,1\}^n ~|~ s\in S\},\\
\gamma_B(V) &\ud \{s\in \States ~|~ \langle s\models p_1,...,s\models p_n \rangle 
\in V\}.
\end{align*}

\noindent
These functions give rise to a disjunctive 
(i.e., $\gamma_B$ preserves arbitrary lub's in $\ok{\langle\wp(\{0,1\}^n),\subseteq\rangle}$) Galois connection $(\alpha_B, 
\wp(\States)_\subseteq, \wp(\{0,1\}^n)_\subseteq, \gamma_B)$.
 
Verification of reachability properties 
based on predicate abstraction consists in computing the 
least fixpoint  of the best correct approximation of $\post$ on 
the Boolean abstraction $B$, i.e., $\post^B \ud \alpha_B \comp \post \comp \gamma_B$. 
As argued in \cite{bpr03}, the Boolean abstraction $B$ may be too costly for
the purpose of reachability verification, so that one usually 
abstracts $B$ through the so-called \emph{Cartesian abstraction}. 
The Cartesian abstraction 
is defined as $$C\ud \tuple{\{0,1,*\}^n \cup \{\bot_C\},\leq}$$ 
where 
$\leq$ is the component-wise ordering between tuples 
of values in $\{0,1,*\}$ ordered by 
$0 < *$ and $1< *$, while $\bot_C$ is a bottom element that represents
the empty set of states. 
The concretization function
$\gamma_{C} :
C\rightarrow \wp(\States)$ is as follows:
$$\gamma_{C} (\tuple{v_1,...,v_n}) \ud \{s\in \States ~|~
\tuple{s\models p_1,...,s\models p_n} \leq \tuple{v_1,...,v_n}\}.$$
This latter 
abstraction formalizes precisely the abstract $\post$ 
operator computed by the verification
algorithm of the c2bp tool in SLAM~\cite{slam02}. However, the Cartesian
abstraction of $B$ may cause a loss of precision, so that this abstraction
is successively refined by reduced 
disjunctive completion and the so-called focus operation, 
and this formalizes the bebop tool in SLAM~\cite{bpr03}. 

Let us consider the following example program, 
taken from \cite{bpr03}, where the goal
is that of verifying that the assert at line $(*)$ is never reached,
regardless of the context in which $\mathit{foo}()$ is called.

{
\begin{flushleft}
~~~~~~\textbf{int} $x$, $y$, $z$, $w$;

\smallskip
~~~~~~\textbf{void} $\mathit{foo}$() \{ 

~~~~~~~~~~\textbf{do} \{

~~~~~~~~~~~~~~$z := 0$;~ $x:=y$; 

~~~~~~~~~~~~~~\textbf{if} $(w)$ \{ $x$++;~$z:=1$; \}

~~~~~~~~~~\} \textbf{while} $(!(x = y))$

~~~~~~~~~~\textbf{if} $(z)$ 

~~~~~~~~~~~~~~\textbf{assert}$(0)$; ~~\text{//} $(*)$

~~~~~~\}    
\end{flushleft}
}

Ball et al.~\cite{bpr03} consider the following 
set of predicates  $\mathcal{P} \ud 
\{p_1 \equiv (z = 0), p_2 \equiv (x=y)\}$ so that 
the Boolean abstraction is
$B = \wp(\{\tuple{0,0},\tuple{0,1},\tuple{1,0},\tuple{1,1}\})_\subseteq$.  
Clearly, the analysis based on $B$ allows us to conclude 
that line $(*)$ is not reachable. 
This comes as a consequence of the fact that
the least fixpoint computation of the best correct approximation
$\post^B$ for the do-while loop provides as result
$\{ \tuple{0,0}, \tuple{1,1}\}\in B$ because:
\begin{align*}
\varnothing \xrightarrow{z:=0;~ x:=y}\{\tuple{1,1}\}  
\xrightarrow{\textbf{if}(w) \{x\text{++};~z:=1;\}}
\{\tuple{1,1}\} \cup \{\tuple{0,0}\} 
\end{align*}
where, according to a standard approach, 
the Boolean guard of the if conditional statement is simply ignored. 
Hence, at the exit of the do-while loop one can conclude that 
\begin{align*}
\{\tuple{1,1},\tuple{0,0}\}\cap p_2 = \{\tuple{1,1},\tuple{0,0}\}\cap 
\{\tuple{0,1},\tuple{1,1}\} = \{\tuple{1,1}\}
\end{align*}
holds, hence $p_1$ is satisfied, so that 
$z=0$ and therefore line $(*)$ can never be reached.   

Let us characterize the correctness kernel of the Boolean abstraction $B$ in this example. 
Let us define $S_1 \ud z:=0;~ x:=y$ and $S_2 \ud x\text{++};~z:=1$. 
The best correct approximations of $\post_{S_1}$ and 
$\post_{S_2}$ on the abstract domain $B$ 
turn out to be as follows:
\begin{align*}
\alpha_B \comp \post_{S_1} \comp \gamma_B = \Big\{&
\{\tuple{0,0}\} \mapsto \{\tuple{1,1}\}, \{\tuple{0,1}\} \mapsto \{\tuple{1,1}\}, 
\{\tuple{1,0}\} \mapsto \{\tuple{1,1}\},\\[-5pt]
& \{\tuple{1,1}\} \mapsto \{\tuple{1,1}\} \Big\}\\
\alpha_B \comp \post_{S_2} \comp \gamma_B = \Big\{ &
\{\tuple{0,0}\} \mapsto \{\tuple{0,0},\tuple{0,1}\}, \{\tuple{0,1}\} \mapsto \{\tuple{0,0}\}, \\[-5pt]
& \{\tuple{1,0}\} \mapsto \{\tuple{0,0},\tuple{0,1}\}, 
\{\tuple{1,1}\} \mapsto \{\tuple{0,0}\} \Big\}
\end{align*}
where the functions are defined for singletons values in $B$ only, since they are lifted to
the whole $B$ by additivity.   
Thus, we have that 
$\img(\alpha_B \comp \post_{S_1} \comp \gamma_B) = \big\{\{\tuple{1,1}\}\big\}$ and
$\img(\alpha_B \comp \post_{S_2} \comp \gamma_B) = \big\{\{\tuple{0,0},\tuple{0,1}\}, 
\{\tuple{0,0}\} \big\}$ so that
\begin{align*}
\max\big(\big\{V\in B~|~ \alpha_B(\post_{S_1}(\gamma_B(V))) = \{\tuple{1,1}\}\big\}\big) &= 
\big\{\{\tuple{0,0}, \tuple{0,1}, \tuple{1,0}, \tuple{1,1}\}\big\}\\
\max\big(\big\{V\in B~|~ \alpha_B(\post_{S_2}(\gamma_B(V))) = \{\tuple{0,0},\tuple{0,1}\}\big\}\big) 
&=  \big\{\{\tuple{0,0}, \tuple{0,1}, \tuple{1,0}, \tuple{1,1}\}\big\}\\
\max\big(\big\{V\in B~|~ \alpha_B(\post_{S_2}(\gamma_B(V))) = \{\tuple{0,0}\}\big\}\big) 
 &=  \big\{\{\tuple{0,1}, \tuple{1,1}\}\big\}
\end{align*}

\noindent 
Hence, by Theorem~\ref{kernel}, the kernel 
$\mathscr{K}_F(B)$ 
of $B$ for $F\ud\{\post_{S_1}, 
\post_{S_2}\}$ is:
\begin{align*}
\Cl_\cap \big( \Cl_\cup \big( 
\big\{ 
\{\tuple{0,0}\},
\{\tuple{1,1}\}, 
\{\tuple{0,0},\tuple{0,1}\}, \{\tuple{0,1},\tuple{1,1}\},
\{\tuple{0,0},\tuple{0,1},\tuple{1,0},\tuple{1,1}\}
\big\}  
\big) \big)  \\
=\Cl_\cup \big( \big\{  
\{\tuple{0,0}\},
\{ \tuple{0,1}\},
\{ \tuple{1,1}\}, \{\tuple{0,0},\tuple{0,1},\tuple{1,0},\tuple{1,1}\}
\big\} \big) 
\end{align*}
where we observe 
that the set $\{\tuple{0,1}\}$ is obtained as the intersection 
$\{\tuple{0,0},\tuple{0,1}\} \cap \{\tuple{0,1},\tuple{1,1}\}$.
This correctness kernel $\mathscr{K}_F(B)$ 
can be therefore represented 
as $$\tuple{\wp(\{\tuple{0,0},\tuple{0,1},\tuple{1,1}\}) \cup \{\tuple{0,0},
\tuple{0,1}, \tuple{1,0}, \tuple{1,1}\},\subseteq}.$$ 
Thus,
$\mathscr{K}_F(B)$ is a proper abstraction of the Boolean abstraction 
$B$ that, for example, 
is not able to express precisely the property 
$p_1 \wedge \neg p_2 \equiv (z=0) \wedge (x\neq y)$.

It is interesting to compare this correctness kernel $\mathscr{K}_F(B)$ with 
Ball et al.'s~\cite{bpr03} Cartesian abstraction of $B$  defined above. 
It turns out that 
these two abstractions are not comparable. For instance, $\tuple{1,0}\in C$ 
represents $p_1 \wedge \neg p_2$ which is instead not represented 
by $\mathscr{K}_F(B)$, while $\{\tuple{0,0},\tuple{1,1}\}\in \mathscr{K}_F(B)$ 
represents $(\neg p_1 \wedge \neg p_2) \vee (p_1 \wedge p_2)$ which is
not represented in $C$. However, while the correctness kernel guarantees 
no loss of information in analyzing the program $P$ (and therefore
the analysis with $\mathscr{K}_F(B)$ 
concludes that $(*)$ cannot be reached), the analysis of $P$ 
with the Cartesian abstraction $C$ is inconclusive because:
\begin{align*}
\bot_C \xrightarrow{z:=0;~ x:=y} \tuple{1,1}  
\xrightarrow{\textbf{if}(w) \{x\text{++};~z:=1;\}}
\tuple{0,0} \vee_C \tuple{1,1} = \tuple{*,*} 
\end{align*}
where $\gamma_C(\tuple{*,*}) = 
\States$, so that with the abstraction $C$ at the exit of the do-while loop 
one cannot infer that 
line $(*)$ is unreachable.

\section{Related and Future Work}

Few examples of abstraction simplifications are known.  A general
notion of domain simplification and compression in abstract interpretation has been
introduced in \cite{fgr96,gr97} as a formal 
dual of abstraction refinement. This duality
has been further investigated 
in \cite{gm08} to include semantic transforms in a general theory for
transforming abstractions based on abstract interpretation. Our domain transformation
does not fit directly in this framework. Following \cite{gr97}, 
given a property $\mathcal{P}$ of abstract
domains, the so-called core of an abstract domain $A$, when it exists,
provides the most concrete simplification of $A$ that satisfies the
property $\mathcal{P}$, while the so-called 
compressor of $A$, when it exists, 
provides the most abstract simplification of $A$ that induces the same refined abstraction 
in $\mathcal{P}$
as $A$ does.
Examples of compressors include the least
disjuctive basis \cite{gr98}, where $\mathcal{P}$ is the abstract
domain property of being disjunctive, and 
examples of cores include the completeness core
\cite{grs00}, where $\mathcal{P}$ is the domain property of being
complete for some semantic function. The correctness kernel defined in this paper
is neither an instance of a domain core nor an instance of a 
domain compression.
The first because, given an abstraction
$A$, the correctness kernel of $A$ characterizes the most abstract
  domain that induces the same best correct approximation of a
function $f$ on $A$, whilst the notion of domain core for the
domain property $\mathcal{P}^f_A$ of inducing the same b.c.a.\ of $f$ as $A$
would not be meaningful, as this would trivially yield $A$ itself. The second
because there is no (unique) maximal domain refinement of an abstract domain which induces
the same property $\mathcal{P}^f_A$, as shown by Example~\ref{esempio2}. 

The EGAS methodology opens 
some  directions for future work, such as
(1)~the formalization of a precise
relationship between EGAS and CEGAR and (2)~an experimental evaluation
of the integration in the CEGAR loop 
of the EGAS-based refinement strategy of Section~\ref{egas}. 
It is here useful to recall that some work formalizing
CEGAR in abstract interpretation has already been done~\cite{CGR07,gq01,rrt08}. 
On the one hand, 
\cite{gq01} shows that CEGAR corresponds to iteratively compute 
a so-called complete shell \cite{grs00} of the underlying abstract model $A$ 
with respect to the concrete successor transformer, while 
\cite{CGR07,rrt08}  formally compare CEGAR with an abstraction refinement
strategy based on the computations of abstract fixpoints 
in an abstract domain. These works can therefore provide a starting point
for studying the relationship between 
EGAS and CEGAR in a common abstract interpretation setting. 

\paragraph{\textbf{Acknowledgements.}}
This work was carried out during a visit of the authors to
the Equipe ``Abstraction'' lead by P.\ and R.~Cousot,
at \'Ecole Normale Sup\'erieure, Paris. 
This work was partially supported 
by Microsoft Research Software Engineering Innovation Foundation 2013 Award and 
by the University of
Padova under the Projects AVIAMO and 
BECOM.

\end{document}